\documentclass{aa}  

\usepackage{amsmath}
\usepackage{dblfloatfix}

\newcommand{\cm}{cm$^{-1}$}

\newcommand{\txt}[1]{\ensuremath{\mathrm{#1}}}
\newcommand{\EtOH}{\ensuremath{\txt{CH}_3\txt{CH}_2\txt{OH}}}

\newcommand{\rad}{\ensuremath{^\bullet}}
\usepackage{color}

\usepackage{graphicx}
\usepackage{txfonts}
\usepackage{adjustbox}
\usepackage{threeparttable}
\usepackage{ragged2e}
\usepackage{mhchem}
\begin{document}

   \title{UV irradiation of ethanol-containing interstellar ice analogs}

   \subtitle{Photostability in CH$_3$CH$_2$OH:CO mixtures}

   \author{J. A. DeVine\inst{1}\thanks{Main author of this work} \and
            J. Terwisscha van Scheltinga\inst{1} \and
            S. Ioppolo\inst{2} \and
            K.-J. Chuang\inst{1} \and
            E. F. van Dishoeck\inst{3} \and
            T. Lamberts\inst{3,4}
            \thanks{Corresponding author, a.l.m.lamberts@lic.leidenuniv.nl}}

   \institute{
     Laboratory for Astrophysics, Leiden Observatory, Leiden University. P.O. Box 9513, 2300 RA, Leiden, The Netherlands
     \and
     Centre for Interstellar Catalysis, Department of Physics and Astronomy, University of Aarhus, Aarhus 8000, Denmark
     \and
     Leiden Observatory, Leiden University. P.O. Box 9513, 2300 RA, Leiden, The Netherlands
     \and
     Leiden Institute of Chemistry, Gorlaeus Laboratories, Leiden University, PO Box 9502, 2300 RA Leiden, The Netherlands
             }

   \date{Received xx; accepted xx}

 
  \abstract
   {Ethanol (CH$_3$CH$_2$OH) has been detected in interstellar ices within regions associated with the early stages of star and planet formation. Because of its ability to form through multiple gas- and solid-phase pathways, ethanol is found in a range of astronomical environments, including the solid (ice) phase. These diverse conditions can significantly influence its photostability and photochemistry. While laboratory studies have explored the effects of energetic processing on pure ethanol ices, there is still a gap in understanding how ethanol behaves in mixed ices that more accurately simulate astrophysical conditions.}
   {This proof-of-principle study aims to quantify how the composition of the ice influences the photostability of ethanol in CH$_3$CH$_2$OH and CO ice mixtures, from both physical and chemical perspectives. It also seeks to highlight the importance of balancing constructive and destructive processes in shaping the evolution of molecular complexity within interstellar ices.}
   {Ice mixtures with ethanol to CO ratios ranging from 1:0 to 1:11 are exposed to ultraviolet irradiation from a microwave discharge hydrogen lamp under ultrahigh vacuum conditions, simulating the secondary UV photons generated by cosmic rays interacting with hydrogen, at 16~K. The evolution of the solid phase is tracked \textit{in situ} using reflection-absorption infrared spectroscopy (RAIRS), while changes in the gas phase are monitored with a quadrupole mass spectrometer (QMS). Temperature-programmed desorption (TPD) experiments are conducted to aid in the identification of infrared spectral features.}
   {A radiative-transfer model has been developed to more accurately account for the influence of the composition of the ice on the effective photon flux experienced by the molecules embedded within the ice. The model reveals that, during later stages of irradiation, photoproducts play a significant role in the absorbing of incident photons, highlighting the complex cascade of processes initiated by single-photon absorption in ethanol-containing ices. By evaluating photodestruction cross sections as a function of the initial ice composition, we found that CO exerts a stabilizing effect on ethanol. For highly dilute ethanol:CO mixtures—representative of conditions in astronomical ices—the photodestruction cross section of ethanol is estimated to be approximately $1.6\times10^{-17}$ cm$^2$ photon$^{-1}$ after correcting for the effective absorbed UV fluence of the studied interstellar ice analogs.}
   {}


   \maketitle
%

\section{Introduction} \label{sec:intro}

Ice dust grains are now recognized as a critical component in the molecular evolution of interstellar and protostellar regions \citep{Boogert2015}. These ices, mainly composed of simple molecules such as H$_2$O, CO, and CO$_2$, are exposed to a variety of conditions that promote the development of chemical complexity. As a result, they serve as the birthplace of most complex organic molecules (COMs) observed in interstellar space. Understanding the chemical pathways followed by these ice-bound molecules when exposed to energetic processing (e.g., photons, ions, electrons, and thermal energy) is crucial for accurately modeling their observed abundances. Laboratory experiments play a key role in advancing our understanding of these astronomical processes \citep{Cuppen2024}. By recreating space conditions and manipulating factors such as incident ultraviolet (UV) fluence, ice composition, and temperature, these studies provide quantitative and qualitative insight into the chemistry occurring in regions of star and planet formation \citep{MunozCaro2013, Materese2021}.

Methanol (CH$_3$OH), the simplest saturated alcohol, is one of the most abundant molecules found in interstellar ices \citep{Dartois1999,Pontoppidan2004, Boogert2015}. Laboratory experiments investigating the energetic processing of methanol ices have begun to unveil the complexities of interstellar organic chemistry. When exposed to vacuum ultraviolet (VUV) irradiation in the 120-180 nm range—mimicking the secondary photons produced by cosmic-ray interactions with H$_2$ molecules in dense molecular clouds—methanol ices are shown to produce ethanol (CH$_3$CH$_2$OH) and other COMs, signifying an increase in molecular complexity in interstellar ice analogs \citep{Oberg2009,AbouMrad2016,Paardekooper2016,Freitas2020}. These COMs are the result of the dissociation of methanol into radical species, highlighting the delicate balance between constructive and destructive pathways involved in COM formation within interstellar ices.

Solid ethanol (ice) has been detected in various regions of space \citep{Rocha2024,Kou2025,Agundez2023}, and laboratory experiments have demonstrated that it can form from a variety of precursor molecules through numerous mechanisms, many of which involve energetic processing \citep{Bergantini2018,Chuang2021}. From an experimental and theoretical chemistry point of view, it has also been shown that ethanol can form in ices through nonenergetic processes \citep{Bisschop2007,Chuang2020,Perrero2022}. 

Several research groups have explored the chemistry of ice-bound ethanol under energetic conditions, such as X-rays and high-energy protons. For instance, \citet{Hudson2018} irradiated pure alcohol ices, including ethanol, with 1 MeV protons. Reflection-absorption infrared spectroscopy (RAIRS) was used to monitor the conversion of ethanol into acetaldehyde (CH$_3$CHO), which, like other aldehydes and carboxylic acids, exhibits a strong absorption band around 1700~\cm (5.9~\textmu m), distinct from the spectral features of ethanol. Although \citet{Hudson2018} made significant contributions to understanding the relative abundances of alcohols versus aldehydes, particularly for some previously unconsidered larger species, the limited analysis of reaction products raises the question of whether energetic processing of ethanol ices could lead to greater chemical complexity. 

More detailed examination of the reaction products resulting from X-ray and VUV (184~nm) photolysis of ethanol is provided by \citet{Zasimov2023a,Zasimov2023b}. These studies, when ethanol is embedded in noble gas matrices, identify signatures of various radical species and products, including CH$_2$CHOH (vinyl alcohol), H$_2$CCO (ketene), H$_2$CO (formaldehyde), and CH$_4$ (methane), in addition to acetaldehyde, as observed by \citet{Hudson2018}. Several of these species, especially vinyl alcohol and ketene, are highly reactive, although the extent of secondary reactions is limited in the studies of \citet{Zasimov2023a,Zasimov2023b} due to the matrix isolation technique. In addition, the differences in the product distributions between X-ray and VUV photoprocessing highlight the crucial role of excited states of ethanol induced by incident irradiation in determining the final distribution of complex organic molecules.

This work reports experiments designed to investigate how the surrounding environment influences the chemical pathways and overall stability of ethanol in the solid state when exposed to radiation fields typical of interstellar dense molecular clouds. Experiments on pure ethanol ice provided a baseline for the expected chemical complexity, with the observed products consistent with previous findings. Further experiments on CH$_3$CH$_2$OH:CO ice mixtures demonstrate how varying the mixing ratio influences product formation distributions and the photostability of ethanol. The focus on ethanol in CO-rich ices is driven by the widespread abundance of CO in interstellar ices and the fact that key ethanol precursors, such as H$_2$CO and CH$_3$CHO, are formed and embedded in CO-dominated environments \citep{Pontoppidan2006}. Additionally, a proof-of-principle radiative transfer model in ices was applied to quantify ethanol's survivability under astrophysical conditions.

\section{Methods} \label{sec:methods}

\subsection{Experimental methods}
\label{section:series}

Experiments were carried out on the second generation cryogenic product analysis device (CryoPAD$^2$) at the Laboratory of Astrophysics in Leiden \citep{Ligterink2017,Ligterink2018}. This setup is centered on a cryogenically cooled mirror finish gold-plated copper substrate housed in an ultrahigh vacuum (UHV) chamber ($P_{MC}< 5\times10^{-10}$ mbar at $T = 16$~K). A Lakeshore 336 temperature controller maintains surface temperatures in the 16-450~K temperature range, with the sample temperature being read by both a cryogenic silicon diode and a platinum diode. The chemical composition of the ice adsorbed to the gold substrate is monitored through RAIRS, using an externally located Fourier-transform infrared spectrometer (Agilent 660 FTIRS). A quadrupole mass spectrometer (Hiden HAL/3F PIC 1000) is used to simultaneously monitor the composition of the gas phase in the vicinity of the substrate and to capture any desorption processes that may occur. Ices are prepared by passing a gas mixture through a multi-capillary plate controlled by a leak valve into the chamber as described below.

\begin{figure}[t!]
\centering
\includegraphics[width=8cm]{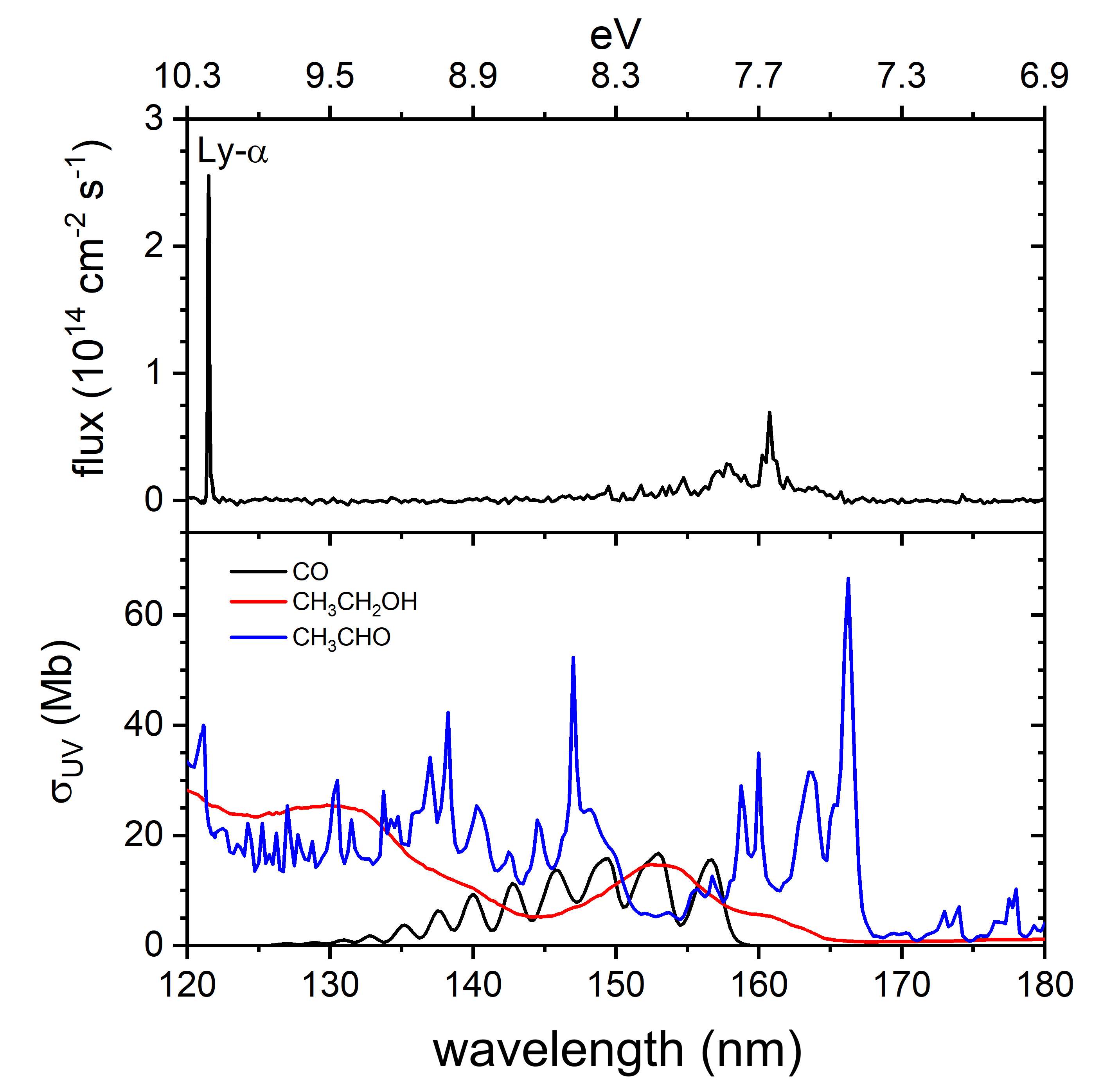}
\caption{(top) Wavelength distribution of the flux of the MDHL used for irradiation. (bottom) VUV absorption cross sections for CO \citep{CruzDiaz2014a} as well as ethanol and CH$_3$CHO \citep{Hrodmarsson2023}  in the lamp region. 1 Mb = 10$^{-18}$ cm$^{2}$. Note that for the two COMs, these correspond to the gas-phase cross sections, and so they should not be quantitatively compared to CO.}
\label{fig:MDHL1}
\end{figure}

\begin{table}[t!]
\centering
\caption{List of CH$_3$CH$_2$OH:CO ice mixtures which were each irradiated for 10 minutes at ca. 16K.}
\def\arraystretch{1.1}
\begin{tabular}{c c | c c c}\hline\hline
& & ethanol & CO & total\\
exp & ethanol:CO & \multicolumn{3}{c}{$10^{15}$ molecules cm$^{-2}$}\\\hline
A & 1:0 & 20 & -- & 20\\
B & 1:2 & 20 & 40 & 60\\
C & 1:3 & 20 & 60 & 80\\
D & 1:6 & 20 & 120 & 140\\
E & 1:8 & 20 & 140 & 160\\
F & 1:0 & 40 & -- & 40\\
G & 1:0 & 60 & -- & 60\\
H & 1:4 & 10 & 40 & 50\\
I & 1:10 & 4 & 40 & 44\\
J & 2:1 & 80 & 40 & 120\\
K & 1:2 & 33 & 47 & 80\\
L & 1:2 & 12 & 22 & 34\\
M & 1:7 & 5 & 35 & 40\\
N & 1:0 & 80 & -- & 80\\
\hline\hline\end{tabular}
\label{table:exps}
\end{table}

To perform energetic processing of the ice samples, a T-type microwave dischage hydrogen lamp (MDHL) is coupled to the main UHV chamber through a magnesium fluoride (MgF$_2$) window. The wavelength-averaged flux of the lamp is measured under typical operating conditions to be $2.5\times10^{14}$~photons~cm$^{-2}$~s$^{-1}$. The wavelength distribution of this flux is measured using a VUV spectrometer (McPherson Model 234/302) situated across the chamber from the lamp inlet. The representative spectrum for the experiments performed in this work is given in Figure~\ref{fig:MDHL1}. More details of the specific operating conditions used can be found in \citet{Ligterink2015} and were chosen to ensure the reproducibility of the spectrum in Fig.~\ref{fig:MDHL1}.

The ice samples studied are summarized in Table~\ref{table:exps}. For each sample, the following protocol was followed. A CH$_3$CH$_2$OH:CO gas mixture reflecting the desired ice composition was prepared on a mixing line that was slightly heated to prevent ethanol from sticking to the stainless steel tubes. The amorphous ices were prepared by passing this mixture through a leak valve into the main chamber, with dosing pressures ranging from $10^{-8}$ to $10^{-6}$ mbar. The ice thickness was estimated from RAIR spectra taken periodically during deposition, and when the desired thickness was reached, the valve was closed. The chamber was allowed to return to base pressure ($<10^{-9}$ mbar) after the end of the dosing, while simultaneously acquiring RAIRS data (average 128 scans per spectrum) and QMS (2-90 amu). The sample was irradiated for ca. 10 minutes, corresponding to a total incident fluence of ca. $2\times10^{17}$ photons cm$^{-2}$. During this time, the RAIRS averaging was changed to 32 scans per spectrum, yielding higher time resolution and noisier spectra. Following irradiation, the sample was left alone for several minutes, with the RAIRS averaging 128 scans per spectrum to allow stabilization of the ice and termination of any potential chain reactions. For samples containing CO in the initial ice, the QMS was intentionally turned off to prevent saturation of the SEM detector and the stabilized ice was subsequently heated from 16 to 40~K at a ramp rate of 3-4~K min$^{-1}$, to allow all CO to desorb from the sample. A temperature-programmed desorption (TPD) experiment was then performed to provide additional information used to assign photoproducts. The sample was heated at a ramp rate of 3-4~K min$^{-1}$ to a terminal temperature of 300-450~K. 

We note that the QMS used in this work is highly sensitive, which allows for detection of weakly contributing species in the gas phase. However, it is unavoidable that QMS-TPD measurements are more susceptible to background desorption than other similar setups, particularly given the temperature gradient between substrate and the rest area of the cold head, which is typically 20-50~K cooler than the sample during warm-up. Thus, species will desorb from the two "substrates" at different times. For this reason, our TPD analysis should primarily focus on temperatures below 200~K, where background desorption from the cold head does not interfere. Any plotted mass signals have been corrected using the calibrated mass correction factor given in the most recent Ph.D. thesis of this set-up \citep{JeroenThesis}.

In designing the experiments summarized in Table \ref{table:exps}, our aim was to acquire several datasets that hold different properties of the initial ice. These are classified as follows
\begin{enumerate}\setlength{\itemsep}{0pt}
    \item Constant initial ethanol column density of ca. $2\times10^{16}$ cm$^{-2}$; vary ethanol:CO mixing ratio and total thickness. [A, B, C, D, E]
    \item Pure ethanol (i.e., constant mixing ratio of 1:0), vary total thickness. [A, F, G, N]
    \item Constant initial CO column density of ca. $3-4\times10^{16}$ cm$^{-2}$; vary ethanol:CO mixing ratio and total thickness. [B, H, I, J, M]
    \item Constant initial total column density of ca. $4-5\times10^{16}$ cm$^{-2}$; vary ethanol:CO mixing ratio (and individual column densities of the components). [F, H, I, L, M]
    \item Constant ethanol:CO mixing ratio of 1:2; vary total thickness. [B, K, L]
\end{enumerate}
By considering these five series of samples, we aim to map out the dependence of ethanol's photostability and photochemistry on factors such as ice thickness and composition. Several experiments were repeated to check for consistency and reproducibility.

\begin{table}
\centering
\begin{adjustbox}{width=\columnwidth}
\begin{threeparttable}
    \caption{RAIRS features observed in irradiated ethanol and ethanol:CO mixtures.}
\begin{tabular}{c c | c c | c | c c c}\hline\hline
\multicolumn{2}{c|}{Peaks (exp.)} & T$_{\rm des}$ (exp.) & amu & species & lit. & $A'$  & ref.\\
(\cm) & (\textmu m) & (K) &  &  & (\cm) & ($10^{-17}$~cm) & \\\hline
2344  & 4.27 & 80 & 44 & CO$_2^a$ & 2343 & 7.6 & [1]\\
*2143 & 4.67 & 30 & 28 & CO & 2139 & 1.1 & [1]\\
2131  & 4.69 & 110 & 42 & H$_2$CCO & 2133 & -- & [2]\\
*1855 & 5.40 & $<30$ & -- & HCO & 1853 & 1.8 & [2]\\
*1843 & 5.43 & $<30$ & -- & HOCO & 1846 & 4.6 & [2]\\
*1726 & 5.80 & -- & -- & H$_2$CO & 1725 & 1.6 & [1]\\
1715  & 5.80 & 110 & 44 & CH$_3$CHO & 1721 & 3.0 & [3]\\
1678  & 5.96 & 200 & 43 & CH$_2$CHOH & 1678 & -- & [4]\\
1636  & 6.11 & 200 & 43 & CH$_2$CHOH & 1639 & 2.8 & [4]\\
*1500 & 6.68 & 110 & 30 & H$_2$CO & 1500 & -- & [1]\\
1428  & 6.70 & -- & -- & CH$_3$CHO & 1428 & -- & [3]\\
1350  & 7.40 & -- & -- & CH$_3$CHO & 1350 & -- & [3]\\
1304  & 7.67 & 50 & 16 & CH$_4^a$ & 1302 & 0.8 & [1]\\
 955   & 10.47 & 140 & 88 & acetoin$^b$ & -- & -- & --\\
 883   & 11.32 & 155 & 31 & ethanol & 886 & 0.3 & [5]\\
\hline
\end{tabular}
\begin{tablenotes}
\raggedright  
\footnotesize
\justifying
\item[]$^a$Unable to be connected to the mechanism in pure samples, due to either a missing product partner or proposed pathway. $^b$Assignment is tentative given the lack of supporting literature.
    References: [1] \citet{Bouilloud2015} [2] \citet{Chuang2020} [3] \citet{Hudson2020} [4] \citet{Bennett2005} [5] \citet{Hudson2017}.
\label{tab:mols}
\end{tablenotes}
\end{threeparttable}
\end{adjustbox}
\end{table}

\subsubsection{Calculation of column densities}
\label{section:cdens}

The column density for a species is obtained from the integrated absorbance $\tilde{A}_{ref}$ of one of its features in the RAIR spectrum via the equation
\begin{equation}
    N = \frac{\ln(10)}{R\times A'}\tilde{A}_{ref}
    \label{eq:cdens}
\end{equation}
where $A'$ is the infrared band strength as determined by transmission infrared measurements and $R$ is a RAIRS correction factor that accounts for the differences in apparent band strengths when operating in reflectance mode. The RAIRS correction factor may not be constant for different molecules, or even different vibrational modes within the same molecule; however, we make a similar assumption as others and take a single value of $R=4.5$ that has been determined in previous experiments in this setup \cite{JeroenThesis}. The band strengths used in this work are given in Table \ref{tab:mols}.

The ethanol column densities are derived from the C-C stretching band at approximately 880~\cm. While spectra obtained during irradiation offer superior time resolution, they are noisier, which could lead to less reliable column densities if these spectra were directly integrated. To mitigate this, the initial ethanol column density is determined by directly integrating the 880~\cm band in the initial spectrum. For subsequent time points, the RAIR spectrum in this region is fitted to an interpolation of the initial spectrum. The scaling factor for each fit is then multiplied by the initial band area and converted to column densities, yielding the final result.

For experiments D and E, the amount of CO was sufficient to saturate the fundamental absorption feature at 2139~\cm. In these cases, the initial column density of CO is estimated from that of the $^{13}$CO feature at ca. 2090~\cm, which has a transmission band strength of 1.7$\times10^{-17}$ cm \citep{Gerakines1995,Bouilloud2015}. The $^{12}$CO column densities are obtained from the $^{13}$CO column density using the natural relative abundances, $N_{12}=93.3~N_{13}$.

For isolated photoproduct features (CO$_2$, H$_2$CCO), the band areas are calculated directly from the spectra, and area uncertainties are obtained by a baseline fitting algorithm. For peaks that overlap with the features of ethanol or other ice components, we extract the peak areas from the adjusted difference spectrum, given by

\begin{equation}\Delta \txt{abs}_t(\omega) = \txt{abs}_t(\omega) - \frac{N_t}{N_0}\txt{abs}_0(\omega)\label{eq:deltaabs}\end{equation}

where $N_t$ and $N_0$ are the ethanol column densities (calculated from the ca. 880~\cm band) at times $t$ and 0, respectively, along with $ \txt{abs_t(\omega)}$ the absorbance at a given frequency. Peak positions and widths were determined by globally fitting lineshapes to the last few spectra, and these parameters were then used to perform individual fits to each peak of interest at each time point in the experiment. In this case, the uncertainties are derived from the confidence intervals of the area parameter in the fit function. Overlapping product peaks (for example, H$_2$CO and CH$_3$CHO in mixed samples) are fitted simultaneously.

\subsubsection{Quantification of photon exposure}\label{section:flucalc}

To zeroth order, the photon exposure of the sample is simply given by the total incident fluence, $\varphi_0\times t_{rad}$, where $\varphi_0=2.5\times10^{14}$ photons cm$^{-2}$ s$^{-1}$ is the photon flux measured for the hydrogen lamp and $t_{rad}$ is the duration of irradiation. However, molecules deeper in the ice will not see the full flux of the H$_2$ lamp. As seen in Figure \ref{fig:MDHL1}, there is non-zero overlap between the hydrogen lamp spectrum and the absorption spectrum of both ethanol and CO, so these ice components will attenuate incident radiation. This attenuation will also change over the course of the experiment, as primary products, such as CH$_3$CHO, are formed.

\begin{table}[b!]
\centering
    \caption{Molecules included in the effective flux calculations described in Section \ref{section:layermodel}.}\label{table:effflux}
\begin{threeparttable}
\begin{tabular}{c | c c | r}\hline\hline
molecule & \multicolumn{2}{c |}{$\sigma(\lambda)$ ref.} & $\tilde{\sigma}_\alpha$ (cm$^2$ molecule$^{-1}$)\\\hline
CO & ice & [1] & $3.9\times10^{-18}$\\
CO$_2$ & ice & [2] & $1.7\times10^{-19}$\\
CH$_4$ & ice & [2] & $2.1\times10^{-18}$\\
\ce{CH3CH2OH} & gas & [3] & $1.0\times10^{-17}$\\
H$_2$CO & gas & [3] & $4.8\times10^{-18}$\\
CH$_3$CHO & gas & [4] & $1.6\times10^{-17}$\\
HCO\rad & gas & [4] & $4.7\times10^{-18}$\\
\hline
\end{tabular}
\begin{tablenotes}
\raggedright  
\footnotesize
\justifying
\item[]References: [1] \citet{CruzDiaz2014a}; [2] \citet{CruzDiaz2014b}; [3] \citet{Heays2017}; [4] \citet{Hrodmarsson2023}.
\label{table:VUV}
\end{tablenotes}
\end{threeparttable}
\end{table}

\begin{table}[b!]\centering
\caption{All reactions following dissociation of the O-H bond in ethanol by a VUV photon. Reaction energies ($\Delta E$) for the ground state are given in units of eV and include zero-point energy corrections}
\def\arraystretch{1.1}
\small
\begin{adjustbox}{width=\columnwidth}
\begin{tabular}{r  rcl | r r}\hline\hline
 &&&&$\Delta E$ (eV)\\\hline
 & CH$_3$CH$_2$O\rad\ + H\rad\ & $\rightarrow$ & \EtOH & -4.24\\
& & $\rightarrow$ & CH$_3$CHO + H$_2$ & -3.64\\
 & CH$_3$CH$_2$O\rad\ + \EtOH\ & $\rightarrow$ & \EtOH\ + CH$_3$CHOH\rad & -0.34\\
 & CH$_3$CHOH\rad\ + H\rad\ & $\rightarrow$ & \EtOH\ & -3.90\\
 & & $\rightarrow$ & CH$_3$CHO + H$_2$ & -3.30\\
 & & $\rightarrow$ & CH$_2$CHOH + H$_2$ & -2.77\\
& CO$_2$ + H\rad\ & $\rightarrow$ & HOCO\rad & -0.11\\
& CO + CH$_3$CH$_2$OH & $\rightarrow$ & HCO\rad\ + CH$_3$CHOH\rad\ & 3.07\\
& CO + H\rad & $\rightarrow$ & HCO\rad\ & -0.83\\
& CH$_3$CH$_2$O\rad\ + HCO\rad & $\rightarrow$ & CH$_3$CHO + H$_2$CO & -2.86\\
& CH$_3$CHO + H\rad\ & $\rightarrow$ & CH$_3$CO\rad\ + H$_2$ & -0.74\\
& CH$_3$CO\rad\ + CH$_3$CHOH\rad\ & $\rightarrow$ & acetoin  & -3.05\\
& CH$_3$CO\rad\ + CH$_3$CO\rad\ & $\rightarrow$ & diacetyl & -2.89\\
 \hline
\end{tabular}
\end{adjustbox}
\label{table:rxncalcs}
\end{table}

For this reason, a model was developed, detailed in Section \ref{section:layermodel} (see Fig. \ref{fig:layerfig}a), which aims to decouple the effects of VUV attenuation from the estimates of photon exposure. This model gives the average flux $\langle\varphi\rangle$ experienced by a molecule in the ice when taking into account the ability of different species to absorb the radiation from the lamp. As shown in Figs. \ref{fig:layerfig}b-c, accounting for the sample composition leads to a decrease in the effective flux and fluence with increasingly thick samples.

From the calculated flux values, we take the time integral to obtain the average fluence,
\begin{equation}\phi_{avg}(t_{rad})=\int_0^{t_{rad}} \langle\varphi\rangle d\tau.\end{equation}
The $\phi_{avg}$ versus time data were fitted to a linear function to obtain the effective flux $\varphi_0^{eff}$, which is given in Table \ref{table:effflux} for each experiment. Using these values in our analysis, we attempt to minimize the influence of ice thickness and composition from the photodestruction cross sections reported in Section \ref{section:xsecs}. Further details on the calculation of photon exposure are provided in Section \ref{section:layermodel}.

The success of this model depends on having both reliable column densities and appropriate photoabsorption data for all ice components. To apply this to our samples, we consider the ice components for which we have assigned RAIRS absorption features and which also have published VUV absorption cross sections (ethanol, CO, CH$_3$CHO, CO$_2$, HCO\rad, CH$_4$, and H$_2$CO; see Table \ref{table:VUV}). Note that radical species (i.e., molecules with one unpaired electron) are indicated by the \rad symbol. With the exception of CO, CO$_2$, and CH$_4$, only gas-phase cross sections are available. To the best of our knowledge, solid-state VUV photoabsorption spectra are not available for the full range of relevant species discussed in this paper. This represents a significant gap in the current literature that warrants further investigation. 

The VUV photoabsorption spectra of molecules exhibit significant differences between the gas and solid phases, primarily due to the influence of intermolecular interactions in ices. In the solid phase, such interactions lead to hypsochromic (blue-shifted) spectral shifts relative to the gas phase, as exemplified by water, where a prominent electronic transition shifts from 167~nm in the gas phase to approximately 144~nm in the condensed phase. Furthermore, while gas-phase spectra typically feature sharp, structured vibrational and rotational bands, solid-phase spectra are broader and often devoid of fine structure due to the suppression of rotational motion and the perturbation of vibrational levels by surrounding molecules. In some cases, additional or modified absorption features arise in the solid phase, which can be attributed to molecular clustering, dimer formation, or structural variations within the ice, such as amorphous versus crystalline arrangements. These effects complicate the assignment of electronic transitions, as condensed-phase transitions often involve delocalized or mixed-orbital characters that are not present in isolated molecules. As a result, theoretical modeling using molecular clusters and fragment-based analysis becomes essential for interpreting solid-phase VUV spectra. 

\subsection{Quantum chemical calculations}

To assess the probability that specific radical-molecule reactions occur in our sample, density functional theory was used to perform geometry optimizations and frequency calculations for a number of relevant molecules at the CAM-B3LYP/def2-tZVP level of theory \citep{CAM-B3LYP}. In the case of the ethanol-assisted isomerization of CH$_3$CH$_2$O\rad~to CH$_3$CHOH\rad, a nudged elastic band calculation was performed to obtain a barrier height of 0.16 eV. All calculations were performed with Orca version 6 \citep{ORCA}. The relevant reaction energies are given in Table \ref{table:rxncalcs}, although these energies are used only as a guide, as they correspond to all reactants / products in the electronic ground state. This selection of reactions is guided by previous work that demonstrated the initial formation of the ethoxy radical upon photolysis of matrix isolated ethanol \citep{Zasimov2023a,Zasimov2023b}. 

\section{Composition-dependent photon exposure in ice mixtures}

\subsection{Layered model for effective flux}
\label{section:layermodel}

The central focus of this work is to investigate how the molecular environment influences the photostability of ethanol within an icy medium. This influence can be broadly attributed to both chemical and physical factors. On the chemical side, neighboring ground-state molecules may serve as reactive partners for photoexcited ethanol, enabling the formation of more stable photoproducts. This dynamic can either accelerate ethanol degradation or promote its regeneration, resulting in a complex and often ambiguous mechanistic landscape where clear trends are difficult to establish. From a physical standpoint, surrounding molecules can attenuate incident VUV radiation, effectively acting as a shielding layer that reduces the photon flux experienced by ethanol molecules embedded within the ice matrix. In this section, we aim to isolate and assess this physical shielding effect by calculating the effective VUV photon flux in our ice mixtures, taking into account the VUV absorption capabilities of the identified ice-bound species.

Consider an ice consisting of an arbitrary number of mixture components $\alpha$, which is subjected to an external incident photon flux whose wavelength dependence is given by
\begin{equation}\mathrm{flux} = \varphi_0 f(\lambda)\;\;\;\mathrm{where}\;\;\;\int_{0}^{\infty}f(\lambda) = 1\end{equation}

\begin{figure}[t!]
\centering
\includegraphics[width=\linewidth]{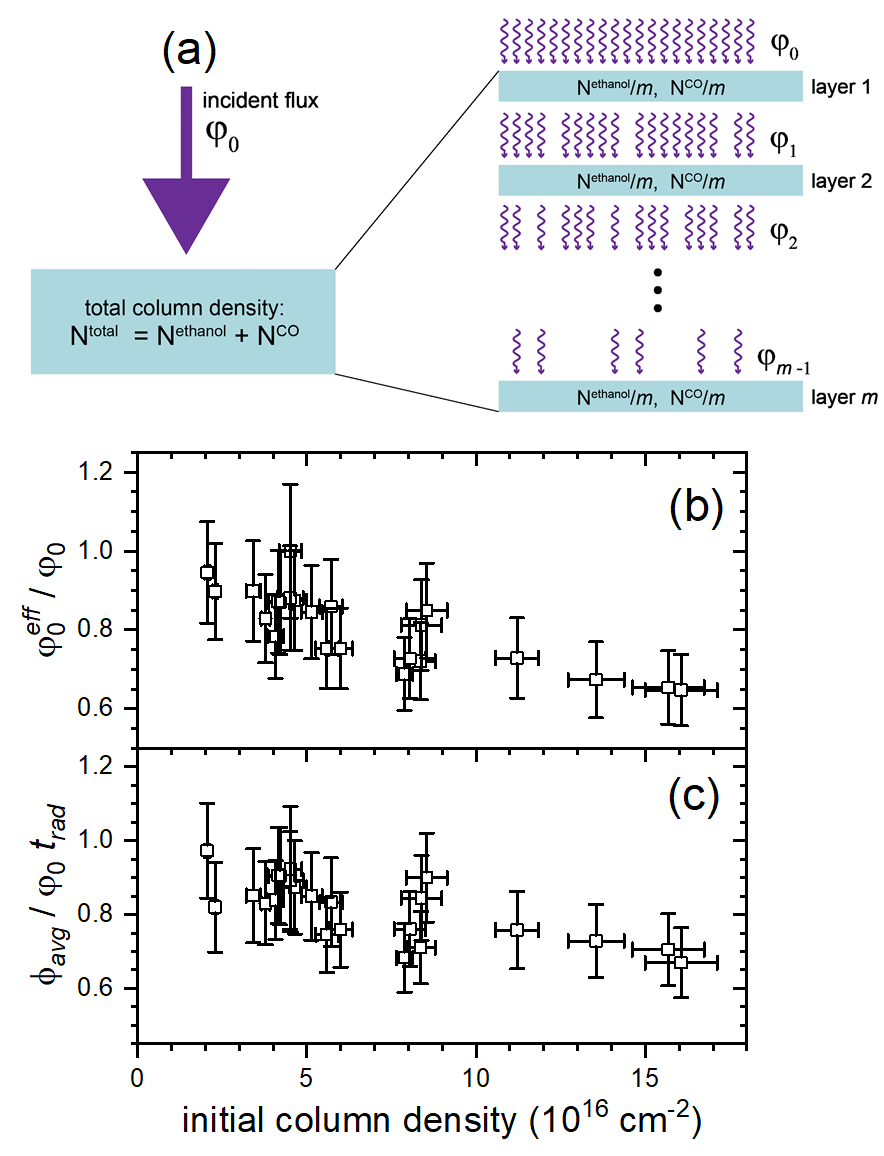}
\caption{An illustration of the layered model of Section \ref{section:layermodel}, used to determine effective photon exposure in mixed ices, is shown in panel (a). In this case, the ice consists of two components, CO and ethanol. The effective flux is considered to be the average of the flux incident on each layer. Panels (b-c) show the result of applying this model to our samples by comparing the ratio of the effective values to the zeroth-order values for the  flux $\varphi_0^{eff}$ and fluence $\phi_{avg}$.}
\label{fig:layerfig}
\end{figure}

To derive this, we consider the ice to be divided into $m$ layers, as shown in Figure \ref{fig:layerfig}a. The topmost layer sees the full incident flux and absorbs some fraction of it; the transmitted flux is then considered the incident flux on the next layer. The effective flux as defined above is given by $\langle\varphi\rangle = \frac{1}{m}\sum_{k=0}^{m-1}\varphi_k;$ with $\psi_k$ (the flux incident on layer $k$) as derived in Appendix \ref{section:app0}, we obtain a closed-form expression for this,
\begin{equation}
    \langle\varphi\rangle = \frac{\varphi_0}{\chi}\left[1-\left(1-\frac{\chi}{m}\right)^m\right]
    \label{eq:avgflux}
\end{equation}
where $\chi=\Sigma_\alpha N_\alpha \tilde{\sigma}_\alpha$ and $\tilde{\sigma}_\alpha$ is the molecule- and lamp-specific effective cross section,
\begin{equation}\tilde{\sigma}_\alpha = \int f(\lambda)\sigma_\alpha(\lambda)d\lambda.\end{equation}
The results of computing $\tilde{\sigma}_\alpha$ using our lamp spectrum are provided in Table \ref{table:VUV} for all molecules included in the effective flux analysis.

The results of this model when applied to our sample can be seen in Figure \ref{fig:layerfig}b; the decrease in the average flux experienced by the molecule can be as large as ca. 40\%. We find that for sufficiently high $\Sigma_\alpha N_\alpha$ (i.e., as long as the ice is sufficiently thick), the choice of $m$ does not significantly impact the results; therefore, for the calculations performed here, we take $m$ to be the total column density $N=\sum_\alpha N_\alpha$ divided by $10^{15}$ molecules cm$^{-2}$.

\begin{figure*}\centering
\includegraphics[width=\linewidth]{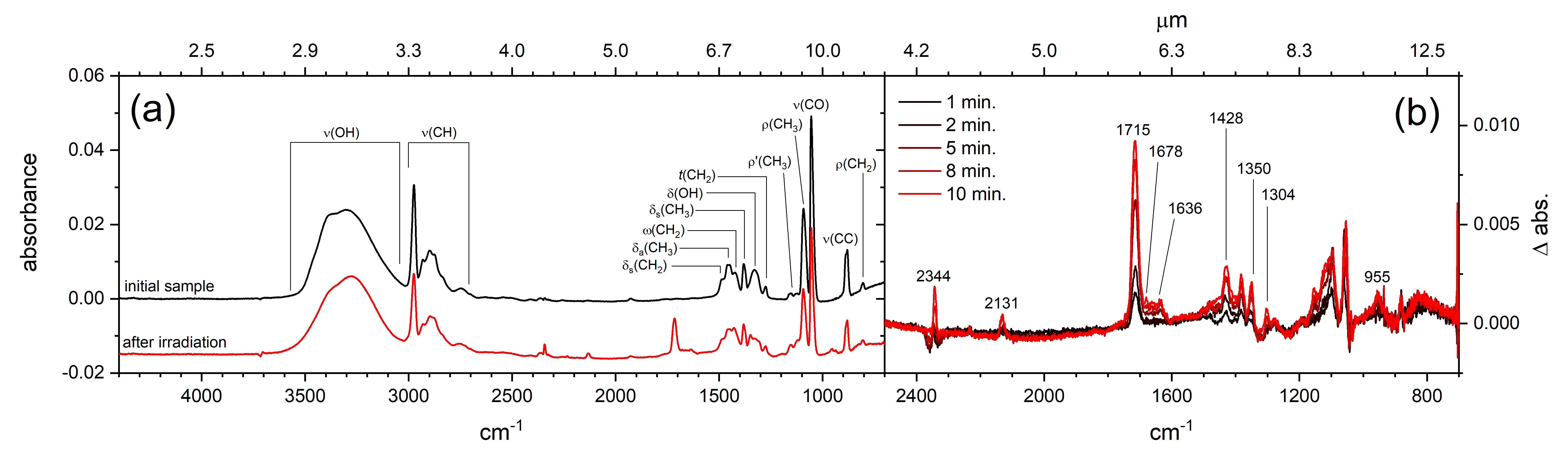}
\caption{(a) RAIR spectra from the initial (black) and final (red) pure ethanol experiments, with the data representing the sum of all pure experiments (A, F, G, N). Peak assignments are taken from \citet{Boudin1998}. (b) Adjusted difference spectra for the sample represented in panel (a) taken at several times during the irradiation.}
\label{fig:purespecs1}
\end{figure*}

\subsection{Distribution of excitation across ice components}\label{sec:partitioning}

In the above flux calculations, we implicitly have some degree of photoexcitation in our samples. When a molecule initially absorbs a photon, it will be promoted to an excited electronic state, after which it may follow numerous pathways to reach stable products. These transformations are rapid (on the order of femtoseconds to picoseconds) and typically involve passing through one or more different reactive states \citep{Andersson2008}. We will consider these species as a whole to represent `reaction centers', and we will refer to the column density of these reaction centers as $N^*$.

The column density of the reaction centers can be considered to be broken down in a similar way to the overall column density, that is, $N^*=\sum N^*_\alpha$, where $N^*_\alpha$ is the column density of the reaction centers originating from photoabsorption by a molecule of type $\alpha$. Not all species represented in the sample absorb the MDHL light equally, and so in general,
\begin{equation}\frac{N^*_\alpha}{\sum_\alpha N^*_\alpha} \neq \frac{N_\alpha}{\sum_\alpha N_\alpha}.\end{equation}
In other words, for each mixture component, its fractional contribution to the overall photochemistry does not reflect its fractional contribution to the sample. This has strong implications for the relative likelihood that different components play a major role in the photochemistry of mixed ices.

To better understand how photon absorption events are divided among the components of our ice mixtures, we make the assumption that the column density of each type of reactive center relies on the effective cross sections $\tilde{\sigma}_\alpha$ as defined in Eq. \ref{eq:chidef},
\begin{equation}
    \frac{N_\alpha^*}{N^*} = \frac{\tilde{\sigma}_\alpha N_\alpha}{\sum_\alpha \tilde{\sigma}_\alpha N_\alpha} = \frac{\tilde{\sigma}_\alpha N_\alpha}{\chi}.
    \label{eq:NstarFrac}
\end{equation}
We refer to this as the partitioning of the reactive centers across the different mixture components, and this can be used to assess the extent to which a particular molecule contributes to the overall photoinduced transformations in our samples.

\subsection{Assessment of multi-photon effects}

The mechanistic interpretation of the photolysis experiments presented herein is critically dependent on the extent to which multi-photon processes contribute under the applied experimental conditions. To quantify this contribution, we employ a layered attenuation model to derive the effective photon flux throughout the ice, as described in Appendix~\ref{section:app1}. This framework enables estimation of the probability that two reactive centers reside in close spatial proximity, sufficient to permit sequential photon absorption events within the same local volume. The resulting analysis demonstrates that such probabilities are vanishingly small unless the timescale for reactive events exceeds 1 ms, a regime that is significantly slower than typical excited-state lifetimes or diffusion-controlled reactions. These findings indicate that, despite the elevated photon fluxes used experimentally relative to astrophysical conditions, multi-photon processes are negligible, thus supporting the relevance of the present results to low-flux environments such as interstellar ices \citep{Cuppen2024}.

On this basis, product assignments are constrained to photochemical pathways initiated by single-photon excitation events. However, the identity of the absorbing species is not fixed to the initial ice constituents. As photolysis proceeds and secondary products accumulate, the absorption profile of the ice evolves, redistributing the incident photon flux among both parent and photoproduct species. This redistribution is analyzed in detail in Section~\ref{sec:partitioning}, where we show that acetaldehyde, a primary photoproduct, becomes a dominant absorber in the later stages of irradiation. Consequently, excitation pathways initiated by CH$_3$CHO must also be considered in the mechanistic assignments, particularly in accounting for late stage photoproducts and spectral changes observed during the course of irradiation.

\section{Photochemistry of ethanol}

\subsection{Pure ethanol}

The initial and final RAIR spectra for the pure ethanol samples subjected to MDHL for an irradiation time of 10 minutes are presented in Fig. \ref{fig:purespecs1}a. Because of the infrared absorption spectrum of pure ethanol and its similarity with those of other molecules possessing similar bonding motifs, the absorption features arising from photoproducts are highly likely to overlap with those of ethanol. To more clearly observe photoproduct peaks, we instead consider adjusted difference spectra (Fig. \ref{fig:purespecs1}b, Eq. \ref{eq:deltaabs}).

In particular, we focus on the fingerprint region between ca. 700-2500~\cm ($\sim$ 4-14 \textmu m). The higher frequency portion of the spectra ($<4$ \textmu m), which includes the OH and CH stretching regions, is highly susceptible to band-shape changes from the local environment, especially given the hydrogen bonding capability of ethanol. In pure samples, as products form, the hydrogen bonding network will be impacted, resulting in changes in this spectral region that do not reflect a specific product species. Although the fingerprint region is not immune to such effects, these structural changes have a smaller impact on the vibrational modes that absorb below 2500~\cm, so our analysis is focused on this spectral region.

The peaks determined to reflect the formation of the product in the adjusted difference spectra are labeled in Fig. \ref{fig:purespecs1}b. In addition to product formation, the peaks in these spectra can reflect morphological and/or electronic state changes in the ice; product formation is asserted only if (1) the peak absorption lies at or near the frequency of a species that (2) can be determined to form following absorption of a single photon by relevant molecules in the ice, considering also that in pure ices the most likely neighbor for any given species is CH$_3$CH$_2$OH. The assignment also relies on the TPD results; the infrared absorption must (3) disappear near to the known desorption temperature of the product in question along with (4) a corresponding desorption event in an appropriate mass channel for this species.

The resultant photoproducts are given in Table \ref{tab:mols}; here, the asterisks indicate the only features observed for mixed samples. Approximate desorption temperatures are provided where possible, as are the dominant mass peaks that are found to correlate with the disappearance of these features. Peaks are correlated with the literature values, and previously reported band strengths are given where available. Additional figures showing the supporting evidence obtained from the TPD measurements are given in Section \ref{sec:TPD} of the Appendix.

\begin{figure}[b!]\centering
\includegraphics[width=7.5cm]{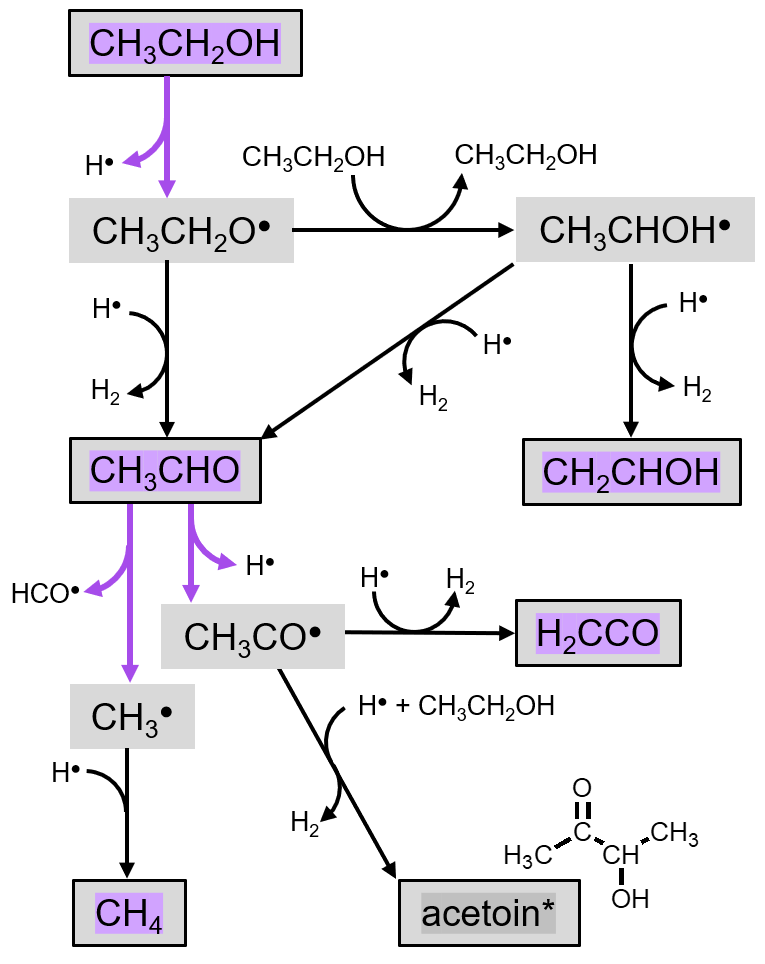}
\caption{Partial reaction diagram reflecting processes in the pure ethanol photolysis experiments. Purple arrows indicate processes induced by photon absorption. Highlighted species have been identified in the RAIR spectra; those with boxes around them are considered to be ``stable,", i.e., do not react further unless exposed to another reactive species. Acetoin is highlighted in gray due to the tentative nature of this assignment.}
\label{fig:rxndiagram}
\end{figure}

We note that because of the short irradiation times (10 minutes), this work contrasts with other studies of energetically processed interstellar ice analogs in that the range of possible processes is somewhat limited. \citet{Jamieson2006} summarized several studies in which pure CO ices were energetically processed; the minimum energy dosage reported was about $5\times10^{18}$ eV cm$^{-2}$, and typical values were 1-2 orders of magnitude higher. In contrast, considering the spectrum of Figure \ref{fig:MDHL1}a, we estimate that the total energy dosage in our samples after 10 minutes of irradiation is
\begin{equation}600\;\txt{s}\times hc\varphi_0 \int\frac{f(\lambda)}{\lambda}d\lambda\approx1.2\times10^{18}\;\txt{eV}\;\txt{cm}^{-2}\end{equation}
where $h$ is Planck's constant and $c$ is the speed of light. In the following, we therefore only consider the processes which occur immediately following single-photon absorption, and we aim to assign these features while developing a mechanistic picture for VUV photolysis of pure ethanol ices. These considerations lead to the partial reaction diagram in Figure \ref{fig:rxndiagram}, which represents the processes identified in pure ethanol.


\begin{figure}[b!]\centering
\includegraphics[width=8cm]{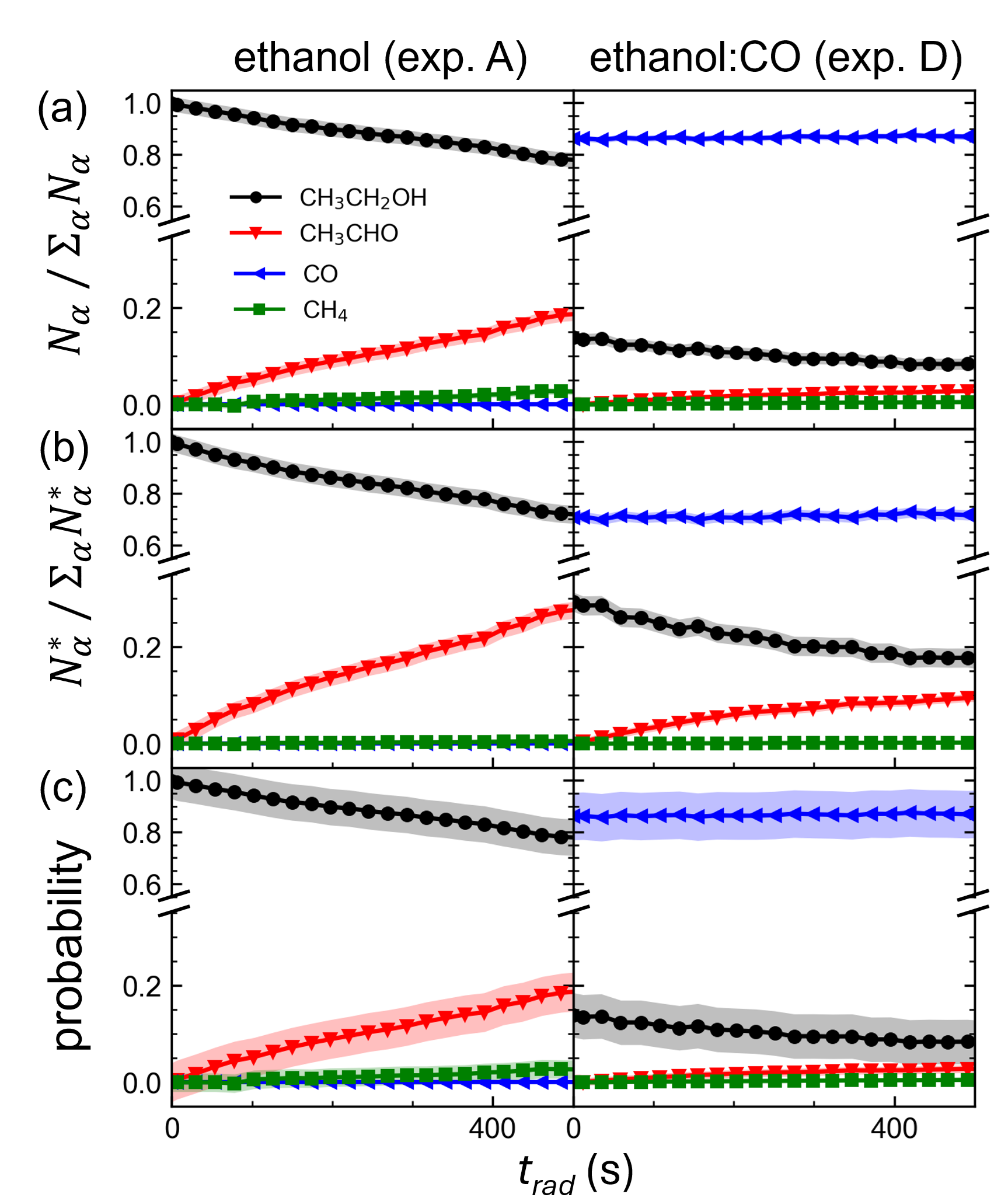}
\caption{Data is provided for a pure ethanol sample (experiment A) as well as an ethanol:CO mixture (experiment D) to illustrate the importance of acetaldehyde as a driver of chemistry in our photolysis experiments. (a) The fractional composition of the sample, over the course of irradiation, as determined from the column densities. Only species with appreciable contributions are shown for clarity. (b) The partitioning of the photon absorption events among the various ice components, as defined in Eq. \ref{eq:NstarFrac}. (c) The probability that an arbitrary molecule in the sample will have the indicated molecule as one of its nearest-neighbors; see Appendix \ref{section:app2}.}
\label{fig:CH3CHOrelevance}
\end{figure}

\paragraph{\underline{H$_2$ loss - CH$_3$CHO and CH$_2$CHOH}:}

Matrix isolation experiments by \citet{Zasimov2023a,Zasimov2023b} have previously shown that upon absorption of a high-energy photon, the primary decomposition pathway for ethanol is loss of the alcohol hydrogen,
\begin{equation}
\txt{CH}_3\txt{CH}_2\txt{OH}+h\nu\rightarrow\txt{CH}_3\txt{CH}_2\txt{O}\rad + \txt{H}\rad.
\label{eq:rxn1}
\end{equation}
The two fragments can react with each other to form acetaldehyde, 
\begin{equation}
    \txt{CH}_3\txt{CH}_2\txt{O}\rad + \txt{H}\rad \rightarrow \txt{CH}_3\txt{CHO}+\txt{H}_2.
    \label{eq:rxn2}
\end{equation}
Indeed, acetaldehyde is one of the first products identified through RAIRS, with the CO stretching band at ca. 1715~\cm appearing almost immediately after opening the VUV shutter. Peaks observed at 1350 and 1428~\cm also correspond to acetaldehyde. This assignment is confirmed by the observation that the feature at ca. 1715~\cm desorbs at around 113~K in the TPD experiments, together with a desorption event in the $m/z = 44$ amu mass channel, consistent with the behavior observed by \citet{Molpeceres2022}. This is shown in Figure \ref{fig:H2lossTPD}, where the relative intensities of the different mass channels are compared to a reference mass spectrum of CH$_3$CHO.

The fragments may also react with a neighboring molecule, in particular ethanol. This is notable as the initial formed ethoxy radical is not the most stable isomer of C$_2$H$_5$O\rad. Our quantum calculations indicate that the ethanol-assisted isomerization,
\begin{equation}
\txt{CH}_3\txt{CH}_2\txt{O}\rad+\EtOH\rightarrow\EtOH + \txt{CH}_3\txt{CHOH}\rad,
    \label{eq:rxn3}
\end{equation}
has a barrier of 0.16 eV and a net energy loss of 0.34 eV. This barrier might seem insurmountable on cryogenic ice, but because the ethoxy radical is likely to initially form in an excited state, this isomerization is expected to be quite facile. Once $\txt{CH}_3\txt{CHOH}\rad$ is formed, we now have access to vinyl alcohol,
\begin{equation}
    \txt{CH}_3\txt{CHOH}\rad + \txt{H}\rad \rightarrow \txt{CH}_2\txt{CHOH} + \txt{H}_2;
    \label{eq:rxn4}
\end{equation}
The formation of CH$_2$CHOH in our samples is signified by the features that appear at ca. 1636 and 1678~\cm. These features disappear in the TPD in the temperature range of 150-200~K alongside a spike in the $m/z = 43$ amu mass channel, in agreement with the behavior observed by \citet{Kleimeier2021} (see Figure \ref{fig:H2lossTPD}).

\paragraph{\underline{Acetaldehyde chemistry - CH$_4$ and H$_2$CCO}:}

As the product column densities develop, it is possible for {these products (CH$_3$CHO and CH$_2$CHOH) to act as drivers of chemistry by absorption of photons. Figure \ref{fig:CH3CHOrelevance}a shows the fractional composition of two samples over the course of irradiation. As detailed in Section \ref{sec:partitioning}, we can use the column densities as well as the effective absorption cross sections in Table \ref{table:VUV} to determine how photon absorption events are distributed among the different components of the sample; this is shown in Fig. \ref{fig:CH3CHOrelevance}b. This shows that towards the end of the irradiation period, acetaldehyde plays a larger role in the attenuation of incident light, indicating that to fully describe the features of the photoproduct in Table \ref{tab:mols} we must also consider reactions initiated by photoexcitation of CH$_3$CHO.

\citet{Zasimov2023c} showed that irradiation of CH$_3$CHO primarily results in dissociation of the C-C bond,
\begin{equation}
    \txt{CH}_3\txt{CHO} + h\nu \rightarrow \txt{CH}_3\rad + \txt{HCO}\rad.
    \label{eq:rxn7}
\end{equation}
The natural subsequent process would be the reaction between these two fragments to form methane and carbon monoxide,
\begin{equation}
    \txt{CH}_3\rad + \txt{HCO}\rad \rightarrow \txt{CH}_4 + \txt{CO}.
    \label{eq:rxn8}
\end{equation}
We see evidence of methane in the pure ethanol samples via its absorption near 1300~\cm; in Figure \ref{fig:methaneTPD}, the behavior of this feature is shown during sample warm-up for experiment G, showing that this peak disappears with low-temperature desorption events around 50~K in the $m/z = 15$ and 16 amu channels. This is consistent with the behavior of methane observed by \citet{Tamai2023}.



 We also see a peak grow at 2131~\cm in pure samples, roughly coincident with the expected absorption of CO at 2139~\cm \citep{Bouilloud2015}; however, this peak does not show the expected desorption behavior for CO, which should desorb between 30-40~K or undergo co-desorption with the dominant ice component, in this case ethanol at 150~K \citep{Burke2008}. Instead, we find that the 2131~\cm feature disappears around 110~K in the TPD, along with a spike in the $m/z = 42$ amu mass channel, as shown in Fig. \ref{fig:keteneTPD}; this is consistent with the behavior previously observed for ketene reported by \citet{Kleimeier2021}. The tentative presence of ketene can be explained by the photochemistry of acetaldehyde. \citet{Zasimov2023c} find loss of H\rad~to be a minor pathway in acetaldehyde photolysis,
\begin{equation}
    \txt{CH}_3\txt{CHO}+h\nu\rightarrow\txt{CH}_3\txt{CO}\rad + \txt{H}\rad,
\end{equation}
leading to the production of ketene (H$_2$CCO),
\begin{equation}
    \txt{CH}_3\txt{CO}\rad+\txt{H}\rad\rightarrow\txt{H}_2\txt{CCO}+\txt{H}_2.
\end{equation}


\begin{figure*}[t!]\centering
\includegraphics[width=\linewidth]{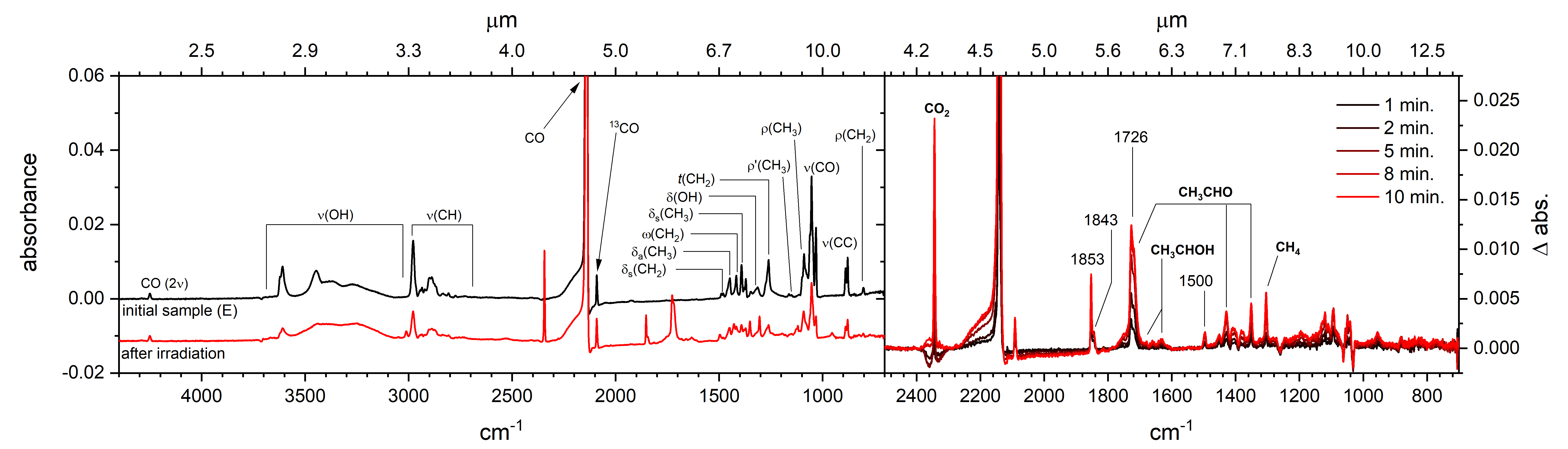}
\caption{(a) RAIR spectra from the initial (black) and final (red) sample E, which was a 1:10 mixture of ethanol:CO. Peak assignments are taken from \citet{Boudin1998}. (b) Adjusted difference spectra for the sample represented in panel (a) taken at several times during the irradiation.}
\label{fig:mixspectra}
\end{figure*}

\paragraph{\underline{Larger COM formation}:}

A complete accounting of all species produced in the irradiated samples is beyond the scope of this work and is particularly difficult given the tendency of larger COMs to have overlapping absorption features. In the remainder of this section we offer some thoughts on possible photoproducts represented in the RAIR spectra that contain more than two carbon atoms.

It is not unlikely that two adjacent CH$_3$CHO molecules will be found in our samples, particularly toward later irradiation times (see Fig. \ref{fig:CH3CHOrelevance}c). Thus, one possible process is
\begin{equation}\txt{H}\rad + \txt{CH}_3\txt{CHO} \rightarrow \txt{H}_2 + \txt{CH}_3\txt{CO}\rad,\end{equation}
which would result in two CH$_3$CO\rad~radicals existing in close proximity. They could then react to form 2,3-butanedione (diacetyl), as previously observed by \citet{Kleimeier2020} in energetically-processed CH$_3$CHO ices. The peak at ca. 955~\cm agrees reasonably well with the 947~\cm absorption feature seen in matrix isolation experiments \citep{Gomez2003}.

It is possible that there are other 4-carbon species reflected in the RAIR spectra; in particular, the broad absorption feature between the 1636 and 1715~\cm bands somewhat resembles the absorptions of glycolaldehyde, which is formed in ices containing methanol \citet{Bennett2007}. An analogous product for ethanol would be 3-hydroxybutanone (acetoin), formed via
\begin{equation}\txt{CH}_3\txt{CHO}+h\nu\rightarrow\txt{CH}_3\txt{CO}\rad + \txt{H}\rad\end{equation}
\begin{equation}\txt{H}\rad+\txt{CH}_3\txt{CH}_2\txt{OH}\rightarrow\txt{CH}_3\txt{CHOH}\rad + \txt{H}_2\end{equation}
\begin{equation}\txt{CH}_3\txt{CO}\rad + \txt{CH}_3\txt{CHOH}\rad \rightarrow \txt{CH}_3\txt{C(O)CH(OH)CH}_3.\end{equation}

Given this pathway and the numerous absorptions of gas-phase acetoin in the relevant spectral region, we consider acetoin as a potential carrier for the 955~\cm feature. While the solid-state infrared spectrum has not been reported in the literature, the fragmentation pattern shows that the high-mass channels may be useful for distinguishing between diacetyl and acetoin. The relevant mass channels are shown in Figure \ref{fig:TPDIR5} and show that the $m/z = 88$ (acetoin) channel agrees reasonably well with the disappearance of the infrared peak, while no clear desorption event is observed for the $m/z = 86$ amu channel (diacetyl). Comparison to the calculated infrared absorption spectrum of gas-phase acetoin shows some agreement with the experiment (Fig. \ref{fig:TPDIR5}), so we consider acetoin to be the most likely candidate.

Another potential COM that may be present in our samples is lactaldehyde, which has been found to form by recombination of HCO\rad~and CH$_3$CHOH\rad~ in the work of \citet{Wang2024}. This species may also contribute to the aforementioned absorption feature bridging the vinyl alcohol and CH$_3$CHO bands near 1700~\cm. A desorption event is observed in the $m/z = 74$ amu mass channel around 100~K, which may correspond to this species.


All candidate species considered in this section exhibit strong absorption features in the vicinity of the CO-stretching mode of acetaldehyde. The potential presence of these species within our samples would therefore result in an overestimation of CH$_3$CHO column densities. This highlights the inherent challenges associated with deconvolving the infrared spectral signatures of complex organic molecules (COMs) in icy matrices and underscores the limitations of infrared spectroscopy as a standalone diagnostic tool for constraining COM formation pathways in such environments.

\subsection{Ethanol:CO}

Figure \ref{fig:mixspectra} shows a figure analogous to Figure \ref{fig:purespecs1} for a 1:10 mixture of ethanol in CO (sample E). Inclusion of CO in the initial ice introduces new reactive pathways via photoexcitation of CO,
\begin{equation}
    \txt{CO}+h\nu\rightarrow\txt{CO}^*.\label{eq:rxn9}
\end{equation}
Even in relatively low-lying excited states of CO, photoexcitation is known to lead to the production of CO$_2$ \citep{DeVine2022},
\begin{equation}
    \txt{CO}^* + \txt{CO} \rightarrow \txt{CO}_2 + \txt{C}.
\end{equation}
This explains the elevated formation of CO$_2$ in the ethanol:CO mixed ices, which absorb at ca. 2344~\cm and disappears during the 80-100~K temperature range during sample warm-up along with a desorption event in the $m/z = 44$ amu mass channel, although the majority of the CO$_2$ co-desorbs with ethanol at ca. 150~K. Note that CO$_2$ is also observed to form in irradiated pure ethanol samples, which may form as one of the products during ethanol irradiation \citep{Wang2025}. Quantification of the CO$_2$ yields is complicated by insufficient purging of the FTIR detector enclosure.

Concurrently to the presence of CO$_2$, we observe another absorption band centered around ca. 1853~\cm, which can be assigned to HOCO\rad. In similar experiments, \citet{Bennett2011} explained the presence of HOCO\rad~through reactions
\begin{equation}
    \txt{CO} + \txt{OH}\rad \rightarrow \txt{HOCO}\rad
\end{equation}
\begin{equation}
    \txt{CO}_2 + \txt{H}\rad \rightarrow \txt{HOCO}\rad.
\end{equation}
This assignment is corroborated by the disappearance of the relatively broad absorption feature at ca. 1843~\cm at around 30~K as shown in Fig. \ref{fig:TPDIR3}, in agreement with the desorption behavior of HOCO\rad~as reported by \citet{Chuang2022}. Overlapping the HOCO\rad~feature is a sharper peak centered around ca. 1853~\cm, which disappears around 20~K; this is consistent with HCO\rad~as noted by \citet{Chuang2022}. This radical can easily form through the addition of the hydrogen radical formed in reaction \ref{eq:rxn1} to a ground-state CO molecule.


The radical HCO\rad~can further react to produce formaldehyde,
\begin{equation}
    \txt{CH}_3\txt{CH}_2\txt{O}\rad + \txt{HCO}\rad \rightarrow \txt{CH}_3\txt{CHO} + \txt{H}_2\txt{CO}.
    \label{rxn:19}
\end{equation}
In the mixed samples, the C=O stretching feature appears to have an additional component beyond the CH$_3$CHO feature seen in the pure samples, which appears at ca. 1726~\cm. Additionally, the mixed samples show a feature that appears at ca. 1500~\cm, which agrees with the observed position of the CH$_2$ scissor mode of \citet{Bouilloud2015}. The C=O stretching frequency of CH$_3$CHO in the pure samples is observed to undergo a blue shift as the sample is heated, complicating the determination of the desorption temperature of the 1726~\cm peak component; however, the 1500~\cm feature shows a behavior consistent with a desorption event in the $m/z = 30$ amu channel (Fig. \ref{fig:formaldehyde}). Thus, the reaction \ref{rxn:19} is likely to take place.

\begin{figure*}[b!]\centering
\includegraphics[width=0.8\linewidth]{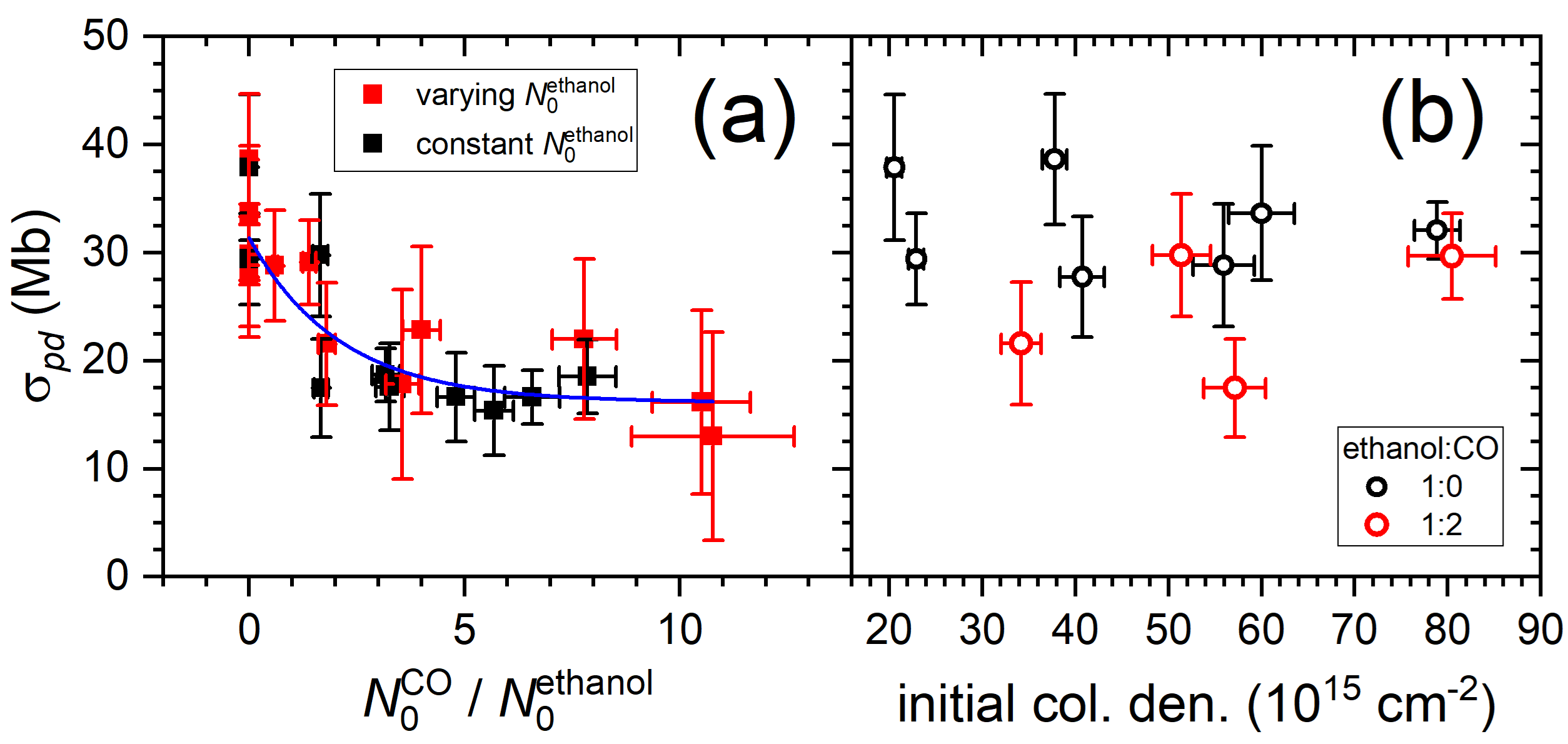}
\caption{Photodestruction cross sections of ethanol in pure and mixed ice samples (1 Mb = $10^{-18}$ cm$^2$). (a) Results are plotted versus the mixing ratio with CO. The black data markers correspond to Series 1, wherein the initial column density of ethanol is fixed at around $2\times10^{16}$ cm$^{-2}$, and the mixing ratio and total initial thickness vary. The blue line shows the fit summarized in Eq. \ref{eq:xsecfit}. (b) Results are plotted for data series where the fractional composition of the initial ice mixture is held constant and the total thickness is varied.}
\label{fig:xsec2}
\end{figure*}

\subsection{Possible fragmentation of the \ce{C-O} bond}
{
Recent theoretical studies of photodissociation of methanol (\ce{CH3OH}) have shown that excitation to higher excited states can lead to the dissociation of the \ce{C-O} bond instead of the \ce{O-H} bond \citep{Cigrang:2024}. Furthermore, theoretical studies in the gas phase also indicate that this could hold true for ethanol, where Rydberg excited state mixing play a key role \citep{Sunanda2019,Barbosa:2022}. 
If the \ce{C-O} bond dissociates, the resulting fragments are the \ce{CH3CH2\rad} and \ce{\rad OH} radicals. The former fragment could result in ethylene and ethane and the latter fragment can result in water or \ce{HOCO}. 
However, given that in bulk ice neither fragment will be able to diffuse effectively, the simple recombination of the two radicals is also possible.\\  
In our experiments we cannot confirm the unequivocal appearance of IR peaks attributed to \ce{C2H4} (949 and 1436 cm$^{-1}$) or \ce{C2H6} (1463 and 2880 cm$^{-1}$) \citep{Moore:1998,Hudson:2014}, indicating that if these molecules are formed, the total amount is limited. The identification of such low amounts in qualdrupole mass spectra is complicated by the fact that the main fragmentation peak for both is expected at m/z = 28, which overlaps with CO. This is a main component of several experiments, but even in the pure ethanol experiments, unfortunately CO is steadily present as contamination in the chamber, not allowing us to detect (low) amounts of \ce{C2H_n}. For the same reason, small amounts of produced \ce{H2O} cannot be detected at m/z = 18. 
While we cannot exclude the formation of species as a result of \ce{C-O} in our experiments, we can state that the overall effect of this dissociation channel is not important under our conditions. It is a clear possibility that while the \ce{C-O} bond might have dissociated in pure or mixed ethanol ices, however, its recombination in the bulk of these ices likely leads to a net zero result.
}

\section{Photostability of ethanol}\label{section:xsecs}
\subsection{Definition and calculation}

The photostability of ethanol in each ice sample is estimated considering the dependence of the RAIRS column density on fluence. Toward the beginning of the experiment, this follows the functional form
\begin{equation}N=N_0 e^{-\sigma_{pd} \phi}\end{equation}
where $N_0$ is the initial column density of ethanol, $\sigma_{pd}$ is the photodestruction cross section, and $\phi$ is the fluence of the photon. We find that for our samples, using the layered model of Section \ref{section:layermodel}, the effective fluence is well described by $\varphi_0^{eff} t_{rad}$, where $\varphi_0^{eff}$ corresponds to the effective flux values given in Table \ref{table:effflux}, so that we obtain the function
\begin{equation}
    N(t_{rad}) = N_0 e^{-k\;t_{rad}}\;\;\;\txt{where}\;\;\;k=\varphi_0^{eff}\times\sigma_{pd}.
    \label{eq:sigfit}
\end{equation}
For each experiment, the ethanol column density was fit to Eq. \ref{eq:sigfit} using a weighting algorithm which accounts for the uncertainties in the column density values; this gives the parameter $k$, which alongside our $\varphi_0^{eff}$ values, yields the photodestruction cross section $\sigma_{pd}$. These values are given in Table \ref{table:effflux} for all experiments.

\subsection{Observed trends and astrophysical implication}

All photodestruction cross sections are plotted as a function of ethanol:CO mixing ratio in Fig. \ref{fig:xsec2}a. Some trends can be identified, particularly for the first series of experiments in which the initial amount of ethanol is kept constant. In these samples, we find that increasing the mixing ratio of CO in the initial ice leads to a noticeable decrease in the photodestruction cross section, indicating that CO has a stabilizing effect on ethanol with regard to VUV irradiation. The other parameters held constant in the five series detailed in Section \ref{sec:methods} did not exhibit significant correlations with ethanol photostability.

This effect can be quantified by fitting all the data in Fig. \ref{fig:xsec2}a to an exponential function,
\begin{equation}\sigma_{pd}(r) = (\sigma_0 - \sigma_\infty) e^{-r/\rho} + \sigma_\infty
\label{eq:xsecfit}
\end{equation}
\begin{equation}\begin{array}{c c l}
\sigma_0 & = & (3.14\pm0.13)\times10^{-17}\;\txt{cm}^2\\
\sigma_\infty & = & (1.61\pm0.17)\times10^{-17}\;\txt{cm}^2\\
\rho & = & 0.47\pm0.17\\
\end{array}\end{equation}
where $r = N_\txt{CO} / N_\txt{ethanol}$ is the mixing ratio of ethanol in CO and $\sigma_0$ is taken as the average of all pure-ethanol $\sigma_{pd}$ values. This fit is shown as a blue line in Figure \ref{fig:xsec2}a. In an astronomical setting, ethanol is expected to be a minor component in CO ices, so the appropriate photodestruction cross section to use is
\begin{equation}\lim_{r\rightarrow\infty} \sigma_{pd}(r) = \sigma_\infty = 1.6\times10^{-17} \;\txt{cm}^2\;\txt{photon}^{-1}. \label{eq:finalanswer}\end{equation}
Typical VUV fluxes experienced by ice-bound molecules in molecular clouds have been estimated by \cite{CecchiPestellini1992} and \cite{Shen2004} to be on the order of $10^4$ photons cm$^{-2}$ s$^{-1}$; combining this with the photodestruction cross section above gives an estimate of $1.4\times10^{5}$ yr for the half-life of ethanol in such an environment. This is notably lower than similar estimations for other COMs, such as ethylamine (3.6$\times10^7$ years, \cite{Zhang2024}, suggesting that ethanol formed in CO-rich ices in the ISM is efficiently destroyed under dense molecular cloud conditions. In fact, recent JWST observations suggesting that ethanol is diluted in a polar environment (that is, molecules with a high dipole moment), in particular, dominated by H$_{2}$O ice \citep{Rocha2024}. A water-rich environment has been shown to provide effective shielding for COMs \citep{Mate2018}. More work is needed to investigate the role of polar ice in the survivability of solid ethanol in space. In fact, the \ce{H2O} VUV absorption spectrum covers the 120 - 160 nm range \citep{CruzDiaz2014a}, thus shielding a different range of wavelengths than CO. Furthermore, (non)energetic formation mechanisms of ethanol in the solid phase can lead to its continuous formation over the lifetime of interstellar clouds, effectively leading to a steady-state concentration in interstellar ices. This is of particular relevance, as the protostellar envelopes in which solid ethanol has been observed have ages of only several $10^4$ years since the collapse of the dense cloud. Thus, as long as ethanol ice formation is still ongoing during or after collapse of the cloud, one can expect enough to survive to allow ice observations.

Note also that \cite{Chen2024} compared the ice and gas ratios of ethanol with respect to methanol toward protostar B1-c. They found that the \ce{CH3CH2OH} abundance relative to \ce{CH3OH} in ice is 1–2 orders of magnitude higher than in the gas phase. This disparity suggests that ethanol may undergo significant gas-phase reprocessing after sublimation or that observational effects from different spatial distributions contribute to the difference.

A potential cause for the stabilizing effect of CO seen in Fig. \ref{fig:xsec2}a may be that CO (or its photoproduct CO$_2$) acts as a VUV shielder, absorbing enough of the incident radiation leading to a decrease in the effective photon exposure of ethanol. However, this effect should be taken into account in our derivation of the average flux based on the model of Fig. \ref{fig:layerfig}a, which explicitly accounts for absorption of photons by CO (and other ice components). Thus, the observed stabilization is unlikely to arise from a purely physical phenomenon, unless there are strong VUV absorbers that are not included in the effective flux calculation. Furthermore, if this was the main mechanism for stabilization, then samples having the same mixing ratio would show a decrease $\sigma_{pd}$ with increasing sample thickness. In other words, chemistry leading to reformation of molecules may appear as shielding by the occurrence of a steady-state abundance. As can be seen in Fig. \ref{fig:xsec2}b, this is not the case, both for pure ethanol and for CO mixtures.

We note that in the ethanol:CO mixtures, the absorption feature near 1850~\cm indicates that photolysis leads to a steady-state concentration of radical species, most likely HCO\rad~and HOCO\rad. These radicals may be the key to stabilizing ethanol with respect to VUV irradiation, as their continued presence in the sample can lead to more chain reactions and introduce new pathways for the reformation of ethanol. In fact, plotting $\sigma_{pd}$ versus the relative steady-state concentration of these radicals, $N_\txt{ss}^\txt{rad} = N_\txt{ss}^\txt{HCO}+N_\txt{ss}^\txt{HOCO}$, shows that elevated levels of these radicals correlate with a decrease in the photodestruction cross section (see Fig. \ref{fig:xsec3}). However, further work would be required to provide a more detailed explanation for this trend.

\begin{figure}[t!]\centering
\includegraphics[width=\linewidth]{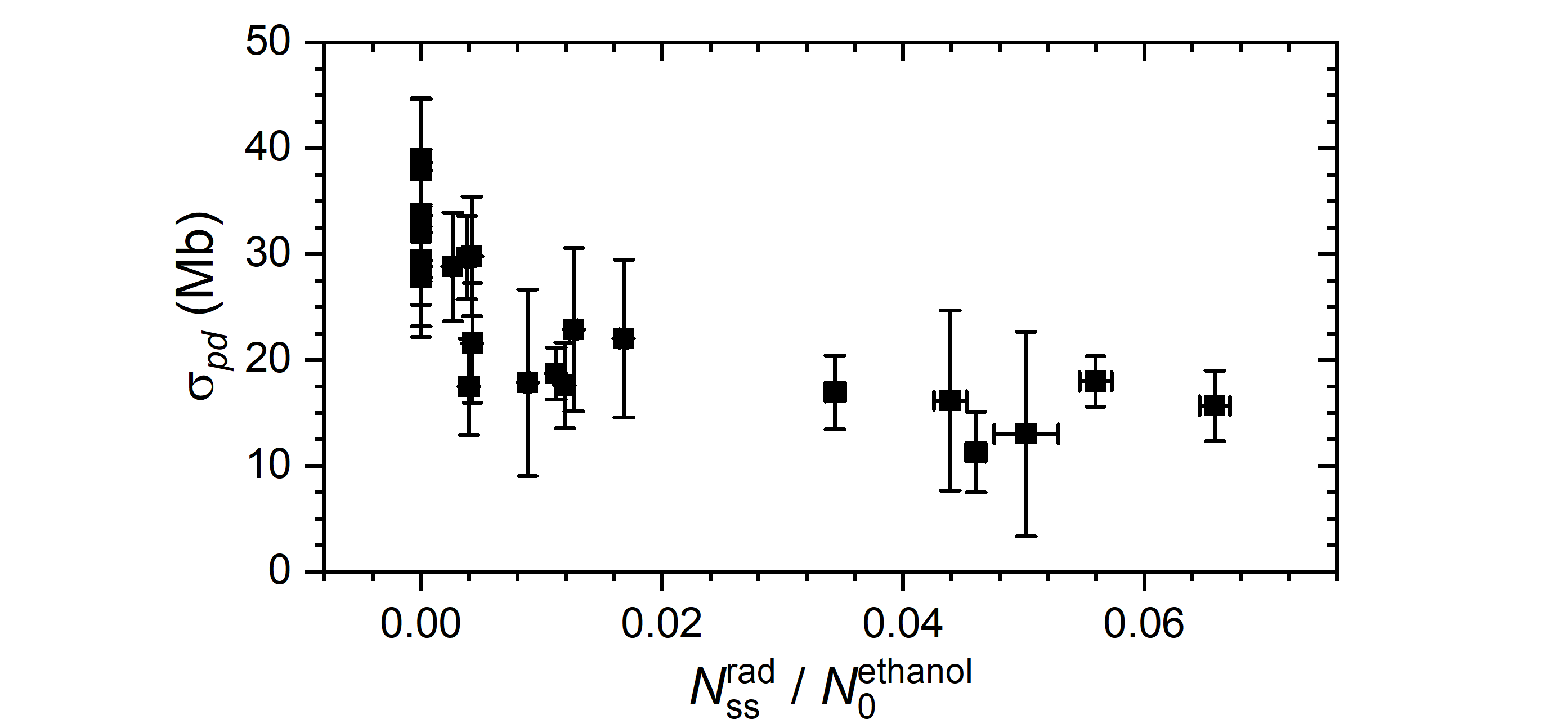}
\caption{The cross sections in Fig. \ref{fig:xsec2}a are plotted versus the steady state concentration of HCO\rad~and HOCO\rad.}
\label{fig:xsec3}
\end{figure}

Finally, although outside the scope of this work, the acquisition of VUV ice cross section datasets across varying temperatures, thicknesses, and morphologies (e.g., amorphous and crystalline phases) is part of an ongoing effort at larger-scale facilities, such as ASTRID2 at Aarhus University, Denmark, with the aim of systematically characterizing environmental effects on spectral features and enhancing the accuracy of molecular identification in astrophysical and planetary observations \citep{Ioppolo2020, Ioppolo2021}.

\section{Conclusion}

The environmental effects on the photostability of ethanol in the 120-180 nm wavelength range have been studied by performing experiments on ethanol:CO ice mixtures of varying mixing ratio. In pure ices, the dominant result of ethanol photoabsorption is hydrogen loss to form acetaldehyde, with vinyl alcohol formation representing a minor channel. Consideration of the VUV absorption cross sections of species identified in the ice shows that further chemical pathways are unlocked by the photoexcitation of acetaldehyde, leading to the production of methane and, potentially, larger COMs.

The irradiation of the ethanol:CO mixtures leads to the formation of the same products, with the addition of formaldehyde, as well as a steady-state concentration of the radical species HCO\rad~and HOCO\rad. A correlation is identified between the photodestruction cross section of ethanol and the steady-state concentration of these species, suggesting a potential chemical explanation for the stabilizing effect that CO has on ethanol. An elevated presence of these radicals enhances the reformation of ethanol. This emphasizes the importance of balancing constructive and destructive pathways when considering the overall stability of COMs in astronomical ices.

Overall, our comprehensive proof-of-principle experiments showed that an ethanol molecule embedded in a CO-rich environment is more stable with respect to VUV irradiation than it is when it is surrounded by other ethanol molecules. However, the destruction of ethanol in CO-rich ices appears to be an efficient process reducing the availability of ethanol over the lifetime of a molecular cloud. The detection of solid ethanol by JWST indicates that it may be embedded in polar environments that can protect it from photodissociation. Efficient solid-phase formation mechanisms will assist in replenishing the amount of ethanol ice observed toward star-forming regions. The models introduced here can be readily applied to other systems, providing a new way to analyze data obtained in laboratory experiments that focus on the energetic processing of interstellar ice analogs.

\begin{acknowledgement}
This paper is dedicated to the late Professor Harold Linnartz.
We thank Ann Mary Wilson for stimulating discussions.
This project has received funding from the Horizon Europe Framework Programme (HORIZON) under the Marie Skłodowska-Curie grant agreement No 101107752 , ``PSI COM''. S.I. acknowledges support from the Danish National Research Foundation through the Centre of Excellence 'InterCat' (grant agreement No. DNRF150).
\end{acknowledgement}

 \bibliographystyle{aa}

\begin{appendix}
\section{Derivation of effective flux}
\label{section:app0}

To derive the effective flux from the layered model of Figure \ref{fig:layerfig}a, we consider the ice to be divided into layers $k=1,2,\cdots,m$, where $m$ is a large integer (see Fig. \ref{fig:layerfig}a). Each layer has the same composition, that is, $N_\alpha/m$ for all components $\alpha$, and the light passes sequentially through each layer. The first layer ($k=1$) is subjected to the full external flux and will absorb some portion of this light according to the Lambert-Beer law; the transmitted (wavelength-averaged) photon flux is given by
\begin{equation}\varphi_{1} = \varphi_0\int_{0}^{\infty}f(\lambda)\left[\prod_\alpha e^{-N_\alpha\sigma_\alpha(\lambda)/m}\right]d\lambda.\end{equation}
This $\varphi_1$ is then considered to be the flux incident on the following layer, $k=2$, which will absorb the same fraction of the light (owing to its identical composition), so the flux transmitted through the second layer is given by
\begin{align*}\varphi_2 &= \varphi_1\int_{0}^{\infty}f(\lambda)\left[\prod_\alpha e^{-N_\alpha\sigma_\alpha(\lambda)/m}\right]d\lambda\\
&= \varphi_0 \left\{\int_{0}^{\infty}f(\lambda)\left[\prod_\alpha e^{-N_\alpha\sigma_\alpha(\lambda)/m}\right]d\lambda\right\}^2.
\end{align*}
Extending this to an arbitrary layer $k$ gives
\begin{equation}
\varphi_k = \varphi_0 \left\{\int_{0}^{\infty}f(\lambda)\left[\prod_\alpha e^{-N_\alpha\sigma_\alpha(\lambda)/m}\right]d\lambda\right\}^k.
\label{eq:app1}
\end{equation}

The average flux incident upon all layers is given by the average of $\varphi_0, \varphi_1, \cdots, \varphi_{m-1}$,
\begin{equation*}
\langle\varphi\rangle = \frac{1}{m}\sum_{k=0}^{m-1}\varphi_k.
\label{eq:app2}
\end{equation*}
If we assume that the components of the ice are all weak absorbers, we can make the approximation
\begin{equation}
\prod_\alpha e^{-N_\alpha\sigma_\alpha(\lambda)/m}=e^{-\sum N_\alpha \sigma_\alpha(\lambda)/m} \approx 1 - \sum N_\alpha \sigma_\alpha(\lambda)/m.
\end{equation}
Defining
\begin{equation}\chi=\sum_\alpha N_\alpha \tilde{\sigma_\alpha}\;\;\;\;\txt{where}\;\;\;\tilde{\sigma}_\alpha = \int f(\lambda) \sigma_\alpha(\lambda) d\lambda,\label{eq:chidef}\end{equation}
we obtain for the average flux the following
\begin{align}
    \langle\varphi\rangle &= \frac{\varphi_0}{m}\sum_{k=0}^{m-1}\left(1-\frac{\chi}{m}\right)^k\\
    &= \frac{\varphi_0}{\chi}\left[1-\left(1-\frac{\chi}{m}\right)^m\right].
\end{align}

\section{Probability of adjacent reaction centers}
\label{section:app1}

\begin{figure*}\centering
\includegraphics[width=10cm]{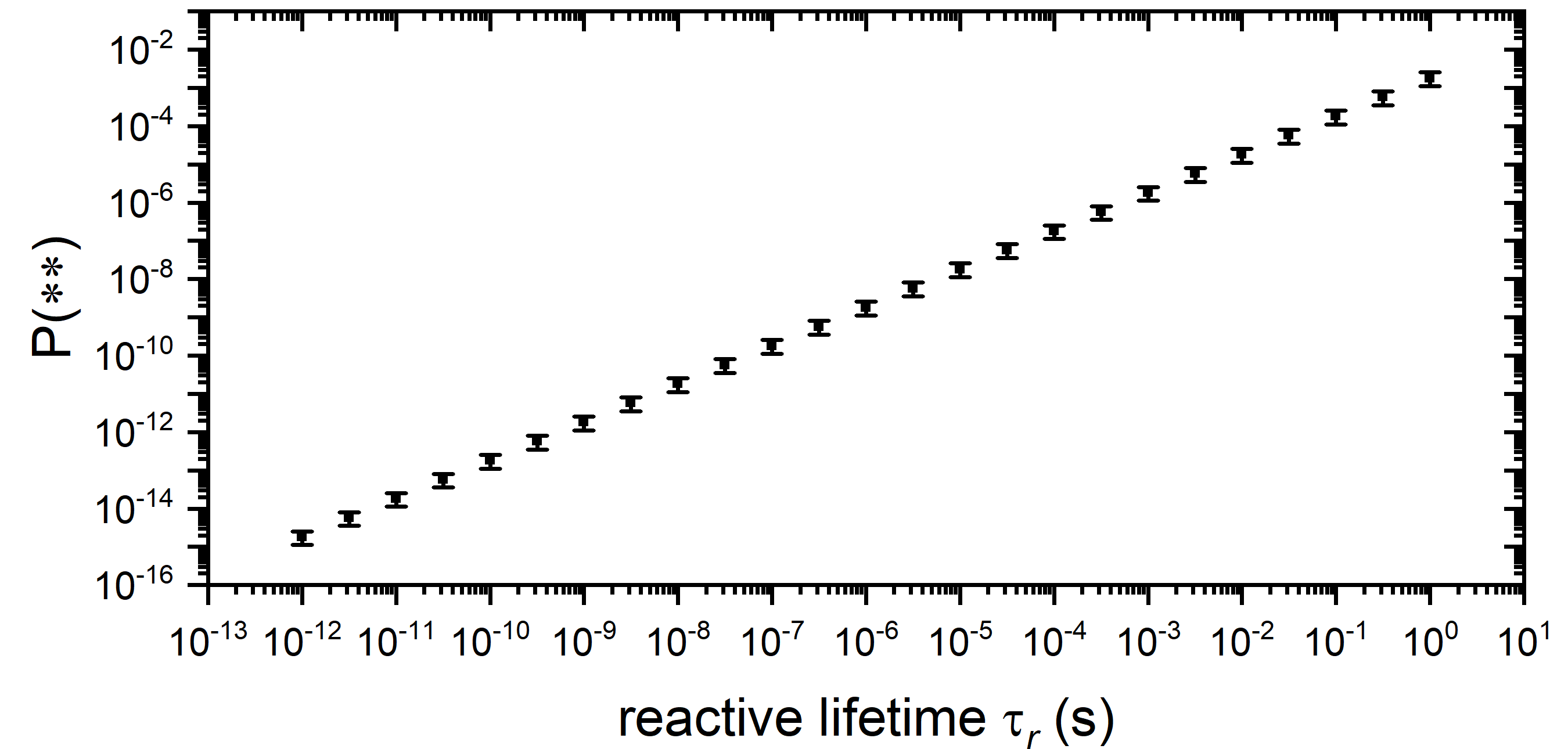}
\caption{Probability that a reaction center will have a nearest-neighbor that is also a reaction center (see Table \ref{table:exps} for sample details), assuming that each molecule in the ice has 10 nearest-neighbors, and plotted as a function of the lifetime of the reactive state. This data is averaged across all samples and all irradiation times.}
\label{fig:app:prob}
\end{figure*}

For the ice samples modeled in Section \ref{section:layermodel}, the change in flux between layers correlates with molecular excitation, so that once irradiation begins, we have some population of reaction centers, $N^*$, corresponding to reactive products of photoexcitation in the sample. Here, we want to determine the probability that two reactive centers would exist in close proximity within our samples. Here, we derive this probability.

Let $\nu$ be the average number of nearest neighbors for a molecule in the sample. We want to determine the probability that, for some excited molecule in the sample, one of its $\nu$ nearest neighbors is ``excited", denoted $P(**)$.\footnote{In this discussion we use the terms ``reactive center" and ``excited molecule" interchangeably for simplicity. Each reactive center is considered to arise from a single-photon absorption event, which brings one of the molecules in the ice to some excited state, which can then fragment or relax or undergo a number of processes. Until it reaches a stable product, representing the dissipation of the absorbed energy, we consider it a reaction center.} To derive this, consider the probability that \emph{none} of the nearest neighbors of the central molecule are excited, which is denoted as $P(!**)$. We obtain this by considering the number of permutations where the $N^*a-1$ excitations are distributed across the $Na-1$ other molecules versus the $Na-1-\nu$ non-neighbor molecules (where $a$ is the area of the sample),
\begin{eqnarray}P(**)=&1-P(!**)
\\=&1-\frac{\left(\begin{array}{c}Na-1-\nu\\N^*a-1\end{array}\right)}{\left(\begin{array}{c}Na-1\\N^*a-1\\\end{array}\right)}\\
=&1-\prod_{k=0}^{\nu-1}\frac{Na-N^*a-k}{Na-1-k}\label{eq:prob}\end{eqnarray}
where
\begin{equation}\left(\begin{array}{c}n\\k\end{array}\right) = \frac{n!}{k!(n-k)!}\end{equation}
is the binomial coefficient, and we have made use of the fact that
\begin{equation}\frac{(x-y)!}{x!} = \prod_{k}^{y-1}\frac{1}{x-k}\iff\frac{x!}{(x-y)!}=\prod_{k}^{y-1}(x-k).\end{equation}

With the probability derived above, we now turn to the determination of $N^*$, the column density of excited molecules in the sample at any given time. Each component will absorb MDHL light differently according to their absorption cross section, $\sigma_\alpha(\lambda),$, where $\lambda$ is the photon wavelength. Assuming that each absorption event is itself a single photon, and using the layered model from Section \ref{section:layermodel}, the total number of molecules excited per unit area per unit time is obtained from the difference in incident and transmitted flux through all $m$ layers,
\begin{equation}
\frac{dN^*}{dt} = \varphi_0 - \varphi_m = \frac{\varphi_0}{\chi}\left[\chi-1+\left(1-\frac{\chi}{m}\right)^m\right]\label{eq:dXdt}
\end{equation}
where $\chi$ is given by Eq. \ref{eq:chidef}. Using VUV cross sections measured in the solid phase (where available) or in the gas phase for each ice component, we can calculate Eq. \ref{eq:dXdt} from our experimental column densities. We find that for all experiments, the number of molecules excited per unit area per unit time is roughly constant throughout the irradiation period, so that $N^*(t)\approx \frac{dN^*}{dt}\times t$.

Now we consider what the relevant values are for $t$. The irradiation time is inappropriate to use because this would imply that there are no relaxation processes for photoexcited molecules. Instead, we consider the "reactive lifetime" of the molecule, loosely defined as the amount of time it takes, following an absorption event, for any photoproducts to be quenched and/or stabilized. As discussed in the main text, for our purposes, this is a more complex quantity than simply the electronic state lifetime, and further will vary between the different components of the ice.

As a first approximation, we assume that there is some characteristic time $\tau_{r}$, which defines the effective lifetime of the reactive centers generated by photoabsorption. This can correlate with, for example, the lifetime of the initial electronically excited state but is considered to be more general. At any given time, the total column density of reaction centers in the sample is given by
\begin{equation}N^* = \tau_{r}\times \frac{dN^*}{dt}.\label{eq:X}\end{equation}

Consideration of the hydrogen bond structure of ethanol indicates that if we limit ourselves to considering only the true nearest neighbors, $\nu<10$ \citep{Jonsson1976}, then this value is used in the plot. The area $a$ is arbitrarily set to avoid the numerical problems that arise when calculating factorials of large numbers.
With Eqs. \ref{eq:prob}, \ref{eq:dXdt}, and \ref{eq:X}, we can calculate from our experimental data the probability that an excited molecule in the sample will have a nearest neighbor that is also excited, which is shown in Figure \ref{fig:app:prob} as a function of $\tau_{ex}$. It is clear from Figure \ref{fig:app:prob} that multi-photon effects are highly unlikely to play any role.

    \section{More general probabilities for molecular pairs}
    \label{section:app2}

    In Appendix \ref{section:app1}, we derived the probability that two excitation centers are nearest neighbors in our sample, which allowed us to rule out the possibility of multi-photon processes. We also provided a way to model the distribution of excitation across components of the sample, which allows us to consider which molecules are most likely to determine the photochemistry in our samples. In this section, we wish to derive a general expression that can be used to determine the probability that a molecule of type $\alpha$ will have a nearest neighbor of type $\beta$. This will allow us to further restrict the processes considered in our mechanism.

The first case we consider is if $\alpha\neq\beta$. In this case, following the procedure used to derive $P(**)$, we consider the number of permutations in which none of the nearest neighbors are of type $\beta$, so we wish to distribute $N_\beta a$ molecules between non-neighbor sites $Na-1-\nu$. This gives a certain level of confidence
\begin{equation}P_\alpha(\beta) = 1-\frac{\left(
\begin{array}{c}Na-1-\nu\\N_\beta a\end{array}
\right)}{\left(
\begin{array}{c}Na-1\\N_\beta a\end{array}
\right)}=1-\prod_{k=0}^{\nu-1}\frac{Na-N_\beta a-1-k}{Na-1-k}\end{equation}
where we use similar arguments as in Appendix \ref{section:app1}, and multiply that result by the probability of finding a molecule $\alpha$ in the sample. In the case $\alpha=\beta$, we have
\begin{equation}P_\alpha(\alpha) = 1-\frac{\left(
\begin{array}{c}Na-1-\nu\\N_\alpha a - 1\end{array}
\right)}{\left(
\begin{array}{c}Na-1\\N_\alpha a - 1\end{array}
\right)}=1-\prod_{k=0}^{\nu-1}\frac{Na-N_\alpha a-k}{Na-1-k}.\end{equation}
The normalization of these probabilities depends on the number of components and nearest neighbors; for the best comparison, we choose to normalize them as follows
\begin{equation}P_{\alpha}^{norm}(\beta)=\frac{1}{\sum_\gamma P_\alpha(\gamma)}\times P_\alpha(\beta),\end{equation}
so that when summed over all species, the probability goes to 1. Note that these probabilities (in the case that $\alpha\neq\beta$) do not depend on the identity of the initial molecule $\alpha$; for example, the probability that an ethanol molecule has a CH$_3$CHO neighbor should be the same as the probability that a methane molecule has a CH$_3$CHO neighbor, that is, $P_\txt{ethanol}(\txt{CH}_3\txt{CHO}) = P_\txt{methane}(\txt{CH}_3\txt{CHO}).$

\newpage
\section{Temperature-Programmed Desorption Data}
\label{sec:TPD}

\begin{figure}[h]\centering
\includegraphics[width=\linewidth]{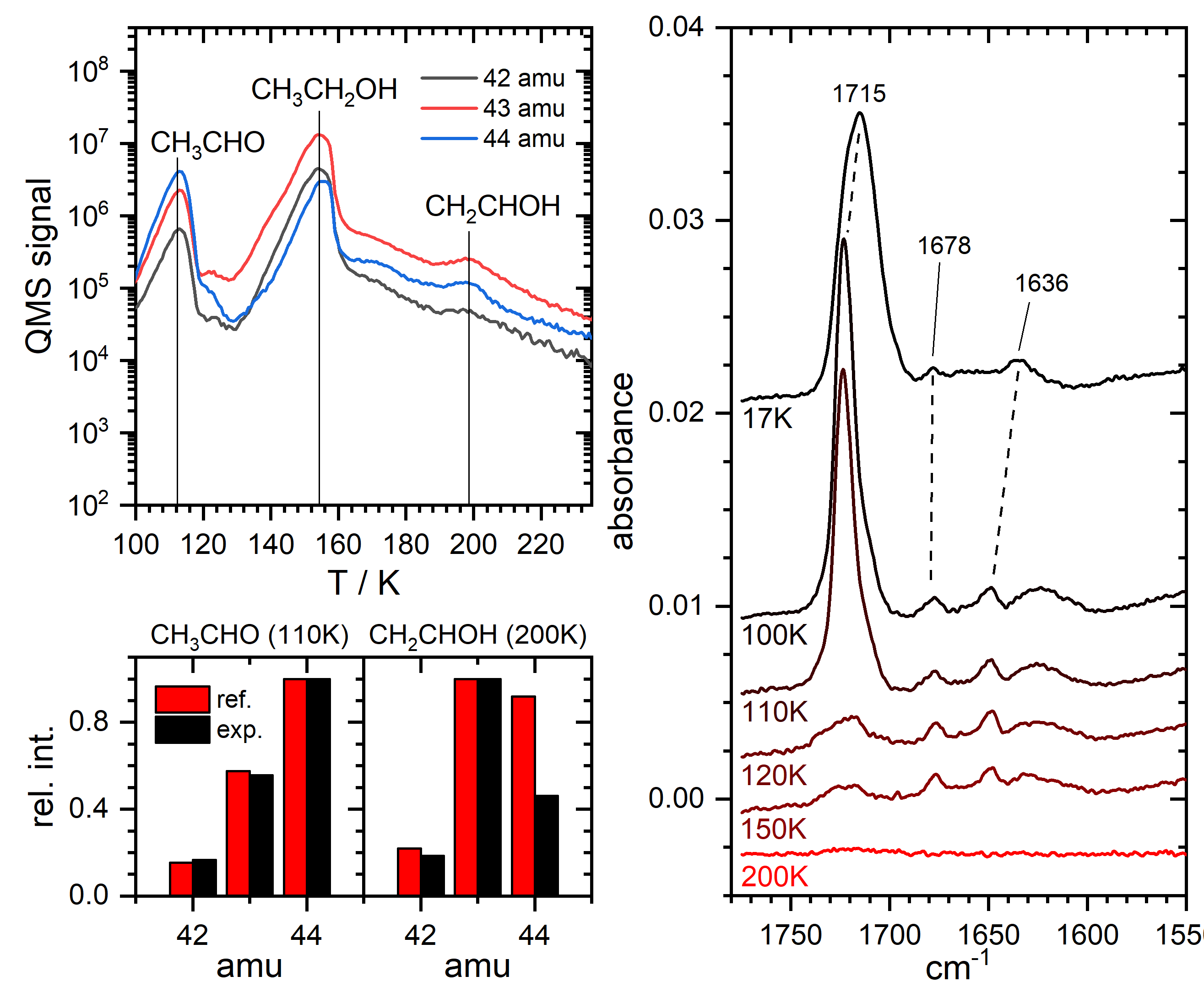}
\caption{TPD signatures of hydrogen loss from ethanol in experiment N. On the left, relevant mass channels for acetaldehyde and vinyl alcohol are shown and compared to reference mass spectra obtained from \citet{Chuang2020}. On the right, RAIR spectra are shown at several temperatures, indicating the changes in peak intensity.}
\label{fig:H2lossTPD}
\end{figure}

\begin{figure}[h]\centering
\includegraphics[width=0.9\linewidth]{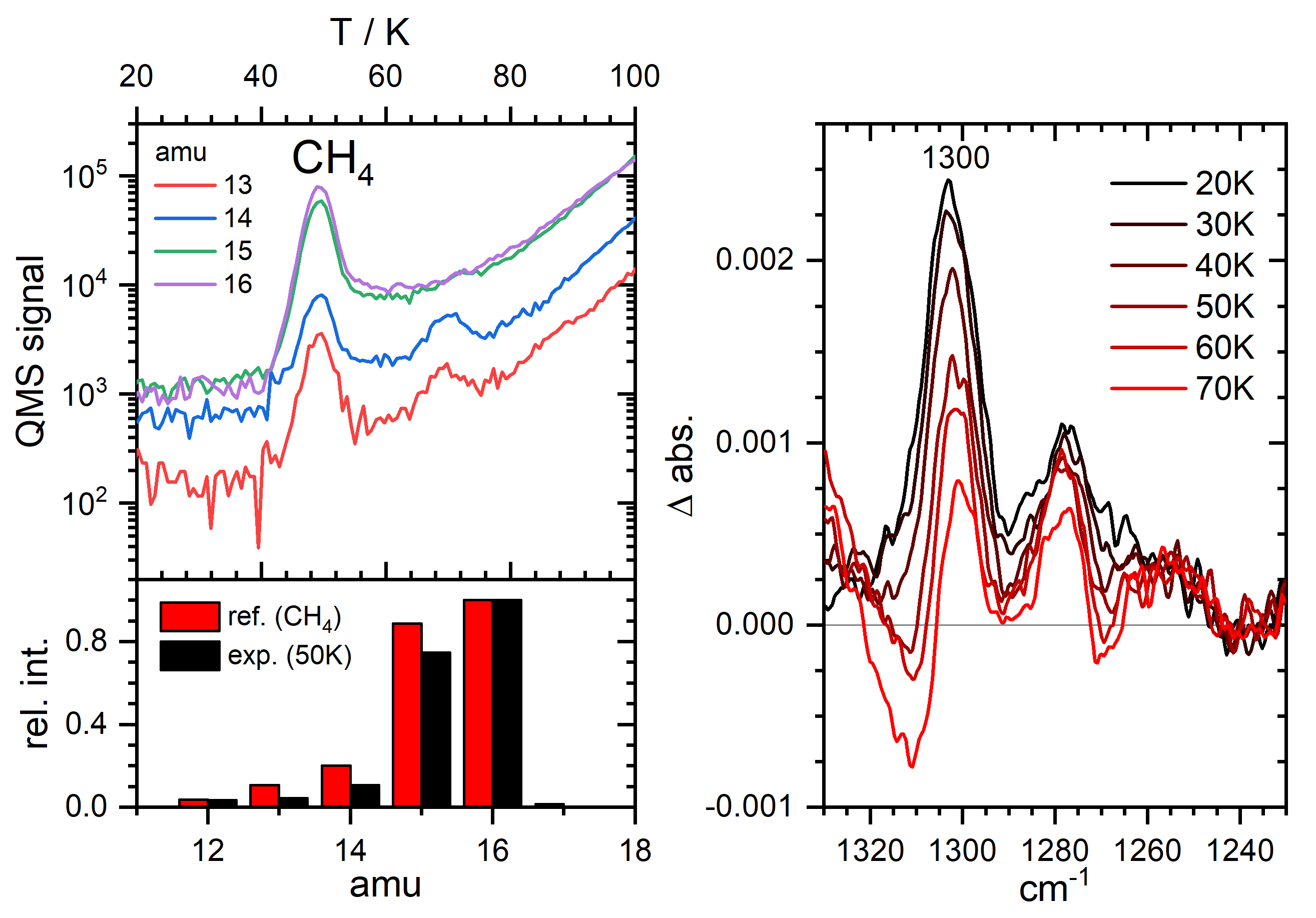}
\caption{TPD signatures of methane formation in experiment G, in which the most CH$_4$ was observed. On the left, relevant mass channels are shown and compared to the reference mass spectrum in the NIST database. On the right, adjusted difference spectra are shown at several temperatures, indicating the changes in peak intensity.}
\label{fig:methaneTPD}
\end{figure}

\begin{figure}[h]\centering
\includegraphics[width=0.9\linewidth]{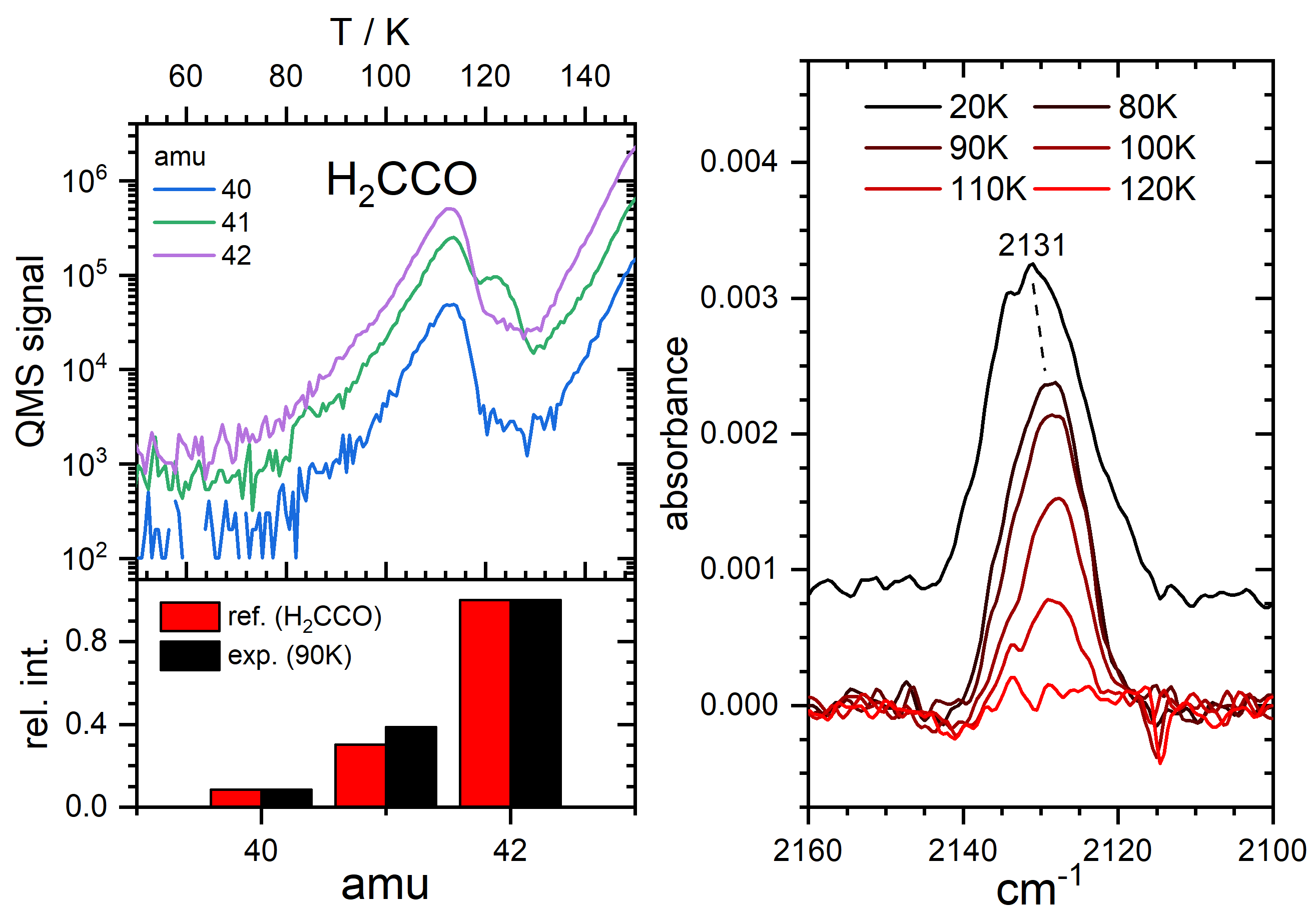}
\caption{TPD signatures of ketene formation in experiment G. On the left, relevant mass channels are shown and compared to the reference mass spectrum in the NIST database. On the right, RAIR spectra are shown at several temperatures, indicating the changes in peak intensity.}
\label{fig:keteneTPD}
\end{figure}

\begin{figure}[h]\centering
\includegraphics[width=\linewidth]{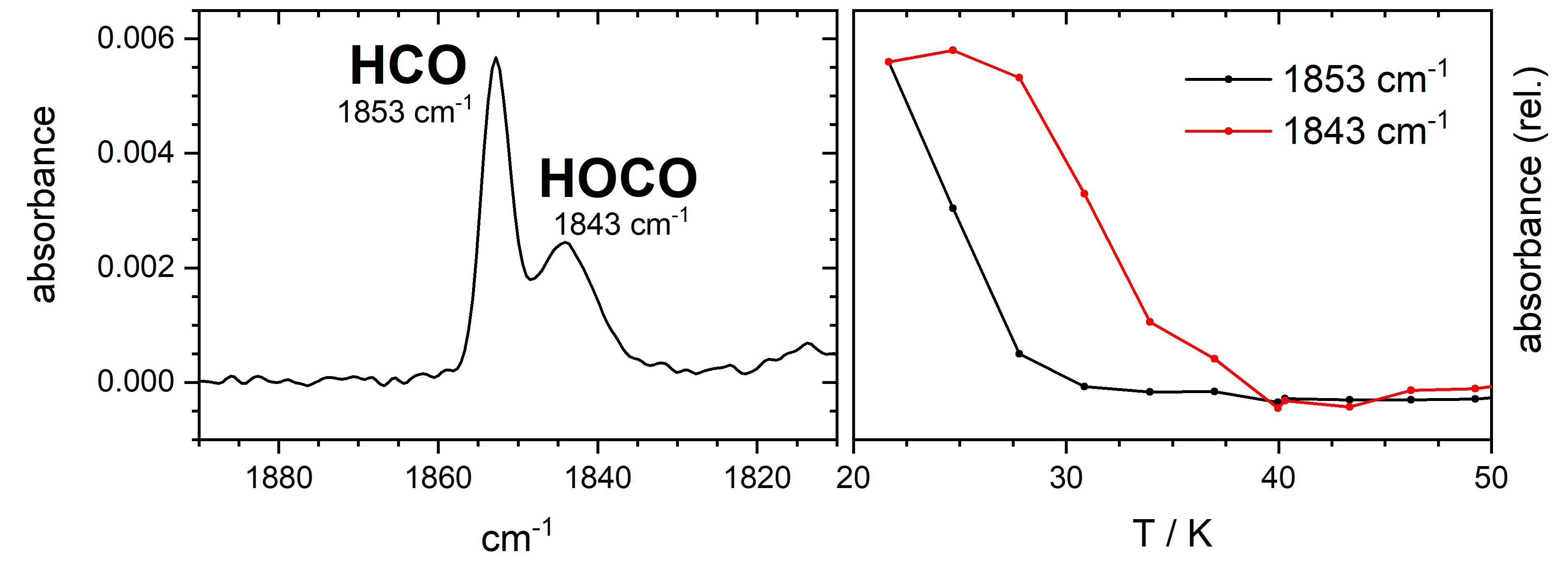}
\caption{(left) Absorption features in sample E assigned to the HOCO\rad~and HCO\rad radicals. (right) Desorption behavior of the peaks shown in the left panel, where we have plotted the absorbance at the indicated frequency. The data is normalized to the initial value to more clearly depict the temperature dependence.}
\label{fig:TPDIR3}
\end{figure}

\begin{figure}[h]\centering
    \includegraphics[width=\linewidth]{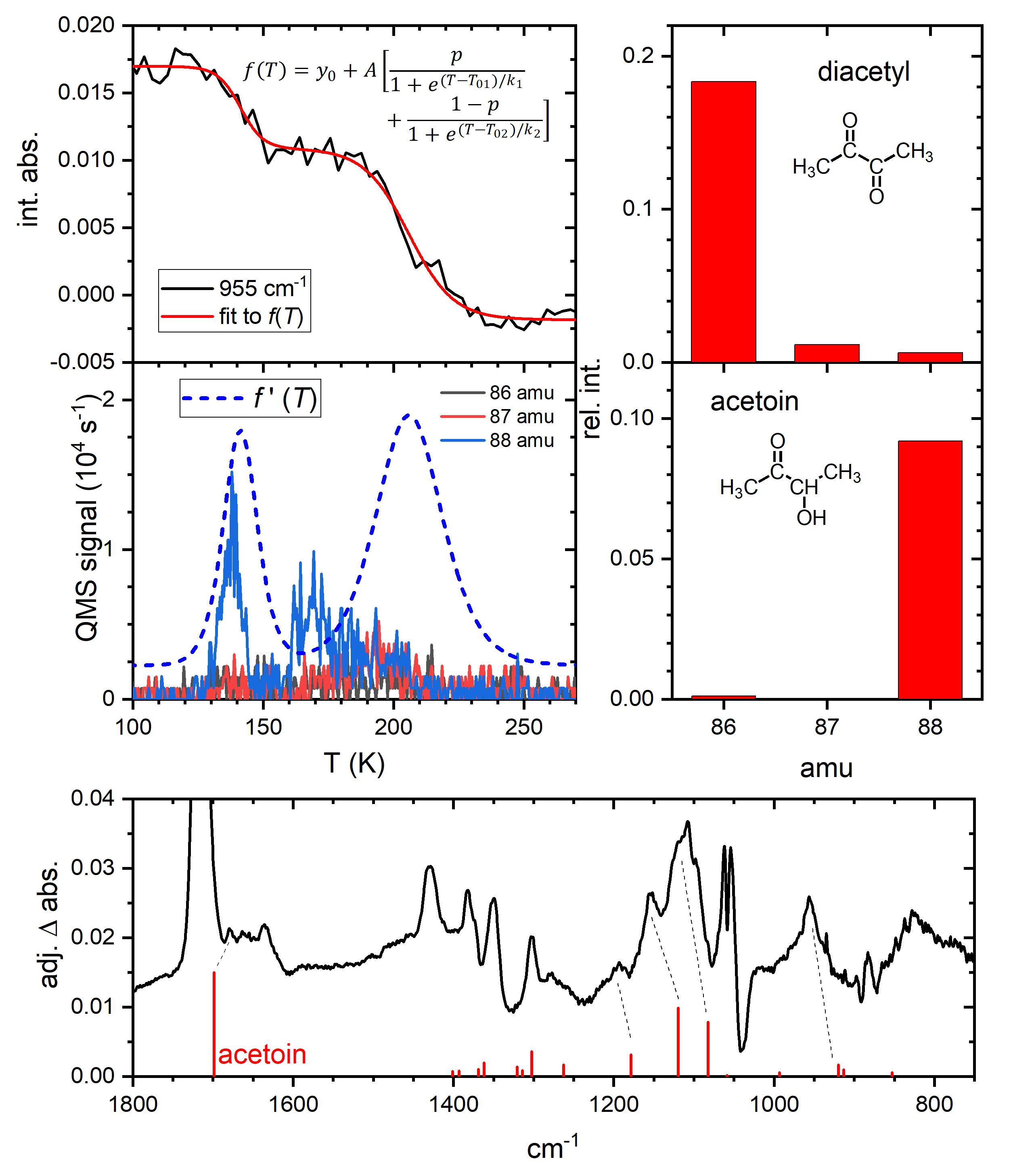}
    \caption{(top-left) The average TPD curve for the ca. 955~\cm feature. These data are fit to the functional form indicated in the inset. (middle-left) The derivative of this fit function is compared to the $m/z = 86-88$ amu mass channels for experiment F. (top- and middle-right) Fragmentation patterns for diacetyl and acetoin, obtained from the NIST database. (bottom) The final adjusted difference spectrum shown in Figure \ref{fig:purespecs1} is compared to the calculated infrared absorption spectrum of acetoin obtained at the CAM-B3LYP/def2-tZVP level of theory. Harmonic frequencies are scaled by a factor of 0.93.}
    \label{fig:TPDIR5}
\end{figure}

\begin{figure}[h]\centering
\includegraphics[width=\linewidth]{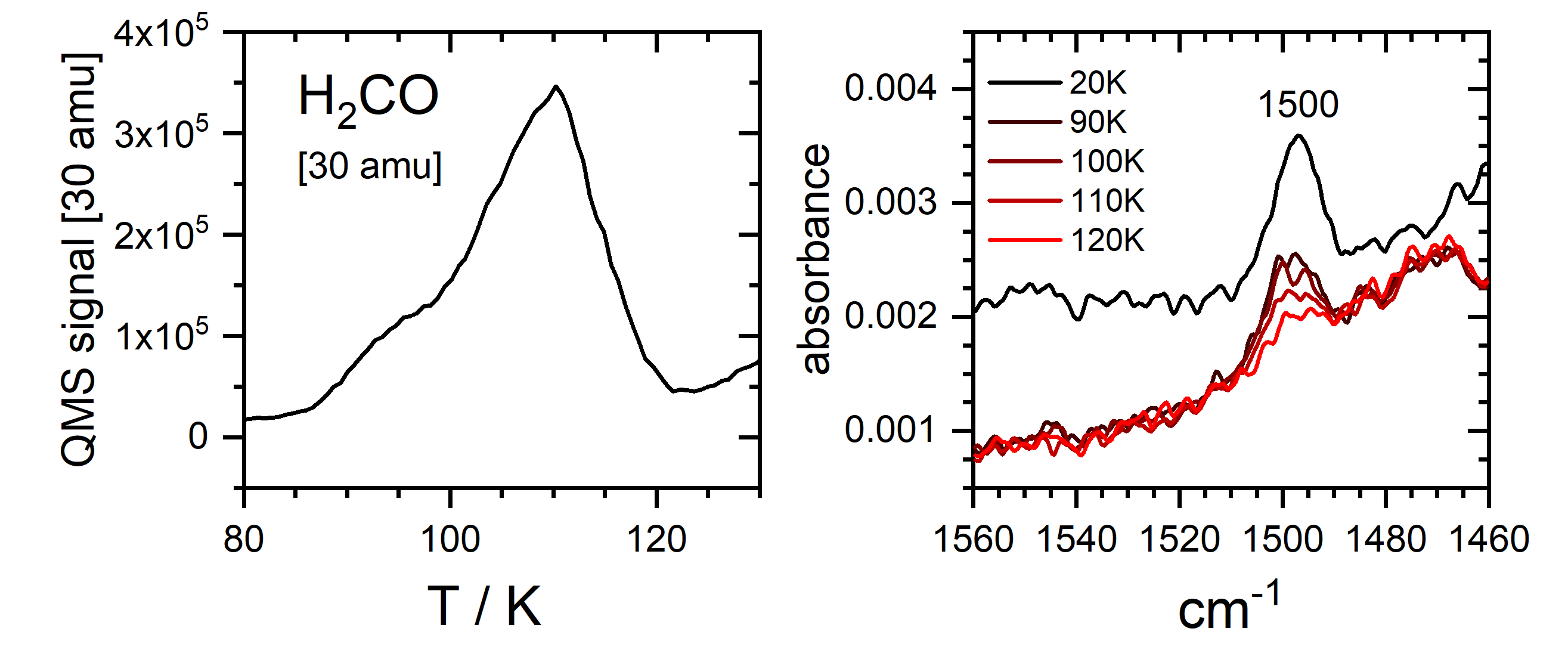}
\caption{TPD signatures of H$_2$CO formation in experiment E. On the left, the QMS signal for the parent mass (30 amu) is shown. On the right, the adjusted difference spectra are shown at several temperatures, indicating the changes in peak intensity which coincides with the ca. 110~K peak in the 30 amu mass channel.}
\label{fig:formaldehyde}
\end{figure}

\end{appendix}

\end{document}